\documentclass[a4paper,fleqn,usenatbib]{mnras}

\usepackage[T1]{fontenc}
\usepackage{ae,aecompl}

\usepackage{graphicx}	
\usepackage{amsmath}	
\usepackage{amssymb}	

\usepackage{newtxtext,newtxmath}

\usepackage{xcolor}

\title[Inhomogenous and anisotropic SSC model]{Effects of inhomogenuity and anisotropy  of radiation field on production and absorption of high energy radiation}

\author[J. Sitarek]{
J. Sitarek,$^{1}$\thanks{E-mail: jsitarek@uni.lodz.pl}
\\
$^{1}$Faculty of Physics and Applied Informatics, University of Lodz,
ul. Pomorska 149/153, 90-236 \L\'od\'z, Poland\\
}

\date{Accepted XXX. Received YYY; in original form ZZZ}

\pubyear{2025}

\begin{document}
\label{firstpage}
\pagerange{\pageref{firstpage}--\pageref{lastpage}}
\maketitle

\begin{abstract}
We investigate the geometrical effects affecting the production and absorption of gamma-ray radiation emitted in inverse Compton scattering in the synchrotron-self-Compton process. 
We evaluate the effect of the anisotropy of the radiation field seen by the scattering electrons homogeneously distributed in the emission region.
Next, we also consider inhomogeneous distribution of electrons and investigate the effect of it in the spherically symmetric emission region.
We also study a cylindrical shape of the emission region and its effect on the isotropy of the emitted radiation. 
We obtain simple numerical factors that scale the emission for different assumptions about the geometry of the emission region and the distribution of the emitting electrons. 
For a 3D Gaussian spatial distribution of the electrons we obtain 0.222 times lower flux than for homogeneous emission region.
Finally, we also evaluate the absorption of the radiation produced in the different scenarios, and compare the full calculations with the two most commonly assumed simplifications. 
We find that for cases when the absorption is lesser than by one order of magnitude, the full calculations for homogeneous sphere can be well approximated with homogeneously-emitting slab, while the absorption in the case of 3D Gaussian distribution of electrons is significantly weaker.
\end{abstract}




\section{Introduction}

Broad-band spectral energy distributions (SEDs) of BL Lac objects are often modelled in terms of a simple Synchrotron-Self-Compton scenario (see e.g. \citealp{ma92,2022Galax..10..105S}).
In order to allow automatic fitting of the spectral energy distribution (SED) to the multiwavelength data, the SSC codes often employ simplifications. 
The open codes computing SSC emission, such as  \texttt{agnpy} \citep{2022A&A...660A..18N} and \texttt{JetSet} \citep{2011ApJ...739...66T,2020ascl.soft09001T} or GAMERA \citep{ha15} typically consider spherical emission regions, however cylindrical regions have been tried as well (see e.g. \citealp{2011MNRAS.416.2368C,2022A&A...658A.173T}).
Inverse Compton and synchrotron emission from spherically-symmetric emission region has been also considered by \cite{1979A&A....76..306G}.
The commonly used assumption that considerably simplifies the calculations of SSC SED is the full homogeneity and isotropy.
Thanks to those simplifications the computations can be done fast enough to allow broad-band fits to the data in order to put constraints on the source internal (e.g.~\citealp{2024ApJ...977....9G,2024MNRAS.535.1484A}) or cosmological (e.g.~\citealp{2011ICRC....8..167A}) parameters.
Not only the electron distribution is assumed to be homogeneous, but also the resulting synchrotron radiation field, which later serves as a target for inverse Compton and $\gamma\gamma\rightarrow e^+e^-$ absorption processes\footnote{It should be noted however that the same codes also often consider directed radiation fields in the case of External Compton calculations rather than SSC.}.
Such an assumption is not self-consistent as even homogeneous electron distribution will result in inhomogeneous and anisotropic radiation field.
The effects of these assumptions on the gamma-ray emission models are even more important in recent years, as such models are discussed in context of constraints on the fundamental physics questions, such as search for Lorentz Invariance Violation (LIV) effects, indirect measurements of Extragalactic Background Light (EBL), or search for Axion-like particles (ALP) (see e.g. \citealp{2022JCAP...02..025L, 2023arXiv231200409A}).

We investigate the effects of the inhomogenuity and anisotropy of the synchrotron radiation field produced by such a simplified homogeneous and isotropic electron distribution.
We then evaluate how this affects the emission and absorption process of the gamma rays. 
We also evaluate the effect of the unknown blob geometry by investigating both (1) a homogeneous, but non-spherically-symmetrical blob and (2) the opposite case, i.e. inhomogeneous, but spherically-symmetrical emission region.


\section{Homogeneous spherical blob model}\label{sec:hom}
As a baseline model, we assume a blob with a radius $R_b$ filled with homogeneous and isotropic electron distribution. 
It also contains a tangled magnetic field with strength $B$. 
Unless stated otherwise, all the quantities are given in the blob's reference frame. 

In the case of blazars, the emission region is moving with a relativistic speed, that affects the observed flux. 
Therefore, in actual applications, one needs to correct for the relativistic beaming (and for far sources also for redshift).
The fluxes and observation angles can be transformed easily between different reference frames using well-known formulae (see e.g. section 5.2.4 in \citealp{mo05}). 
Thus, in order not to introduce additional beaming parameters, we give all the fluxes in the reference frame of the blob. 

We compute the rate of synchrotron differential emission $\dot n_0$ following  Eq.~18 of \citet{fi08} (see also \citealp{cs86}):
\begin{eqnarray}
\dot n_0 &=& \frac{dn_{s}}{d\epsilon dt dV}=\frac{\sqrt{3} e^3 B}{h \epsilon_0 m_e c^2} \int_{\gamma_1}^{\gamma_2} d\gamma n_e(\gamma) R\left(\frac{x_0}{\gamma^2} \right) \\
x_0&=&\frac{4\pi \epsilon_0 m_e^2 c^3}{3 e B h},
\end{eqnarray}  
where $\epsilon_0$ is the energy of synchrotron photons in the units of electron rest mass ($m_e c^2$), 
$n_e(\gamma)=dN_e / d\epsilon dV$ is the differential number density of electrons 
(spreading with Lorentz factor $\gamma$ from $\gamma_1$ to $\gamma_2$),  
$e$ is the elemental charge, 
$c$ the speed of light, 
$h$ is Planck constant and 
$R$ function is defined in Eq.~20 of \citet{fi08}.
We neglect the effect of self-absorption of synchrotron radiation, which is normally relevant only at the lowest energies. 

In the case of homogeneous emission of radiation, the density of the radiation field (differential in terms of energy and direction) will depend linearly on the length of the path along which the radiation field is produced, $x$:
\begin{equation}
n_0(r, \Omega, \epsilon_0) = \frac{dn_s}{d\epsilon dV d\Omega}=\frac{\dot n_0 (\epsilon_0) x}{4\pi c}, \label{eq:n0}
\end{equation}
where division by $4\pi$ is due to division by the full-sphere solid angle. 

\begin{figure}
  \includegraphics[width=0.49\textwidth]{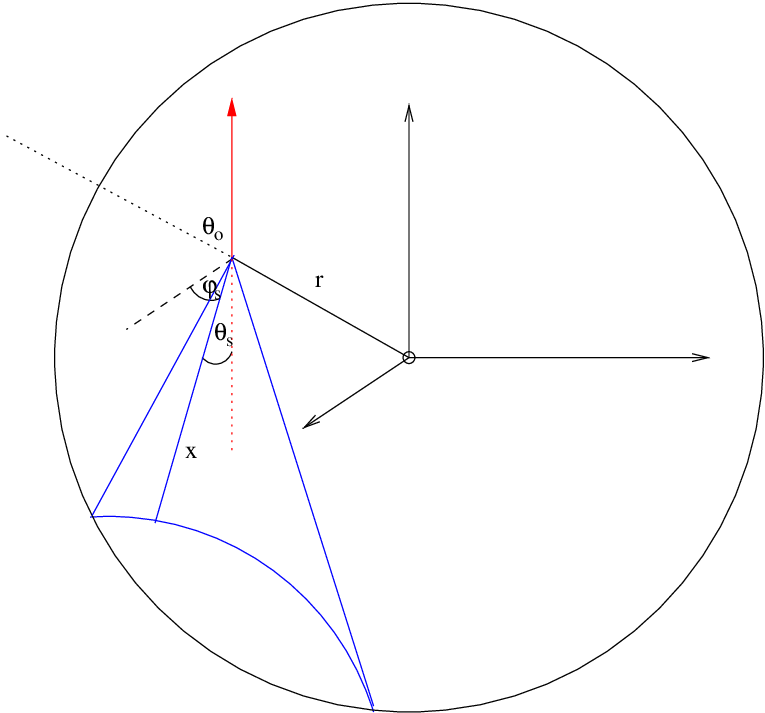}
  \caption{Geometry of radiation field visible by particle at a distance $r$ from the centre of the blob, directed at the angle $\theta_o$ to the observer (red arrow).
    The radiation scattered at an angle $\theta_s$ is located in a blue cone.
  The total length of radiating space along the directions $(\theta_s, \phi_s)$ is marked as $x$.
  The distance of electron from the centre of the blob (measured in the units of blob radius) is $r$.}\label{fig::sketch}
\end{figure}

The distance $x$ can be calculated analytically both inside and outside the blob. 
For a particle oriented at the angle $\theta_o$ to the observer, and at a distance $r \times R_b$ from the blob centre(see Fig.~\ref{fig::sketch}), the radiation originating at the angles $(\theta_s, \phi_s)$ is produced over a path of:
\begin{equation}\label{eq:x}
  x= \left.
  \begin{cases}
    b+\sqrt{d}, & r<1 \\
    2 \sqrt{d}, & r>1\ \mathrm{and}\ d>0\ \mathrm{and}\ b-\sqrt{d}>0\\
    0, & \mathrm{otherwise}    
  \end{cases}
  \right.
\end{equation}
where:
\begin{eqnarray}
  b&=&r(\sin\theta_o\sin\theta_s\sin\phi_s + \cos\theta_o\cos\theta_s) \\
  d&=&b^2 +1 - r^2
\end{eqnarray}

\begin{figure}
  \includegraphics[width=0.49\textwidth]{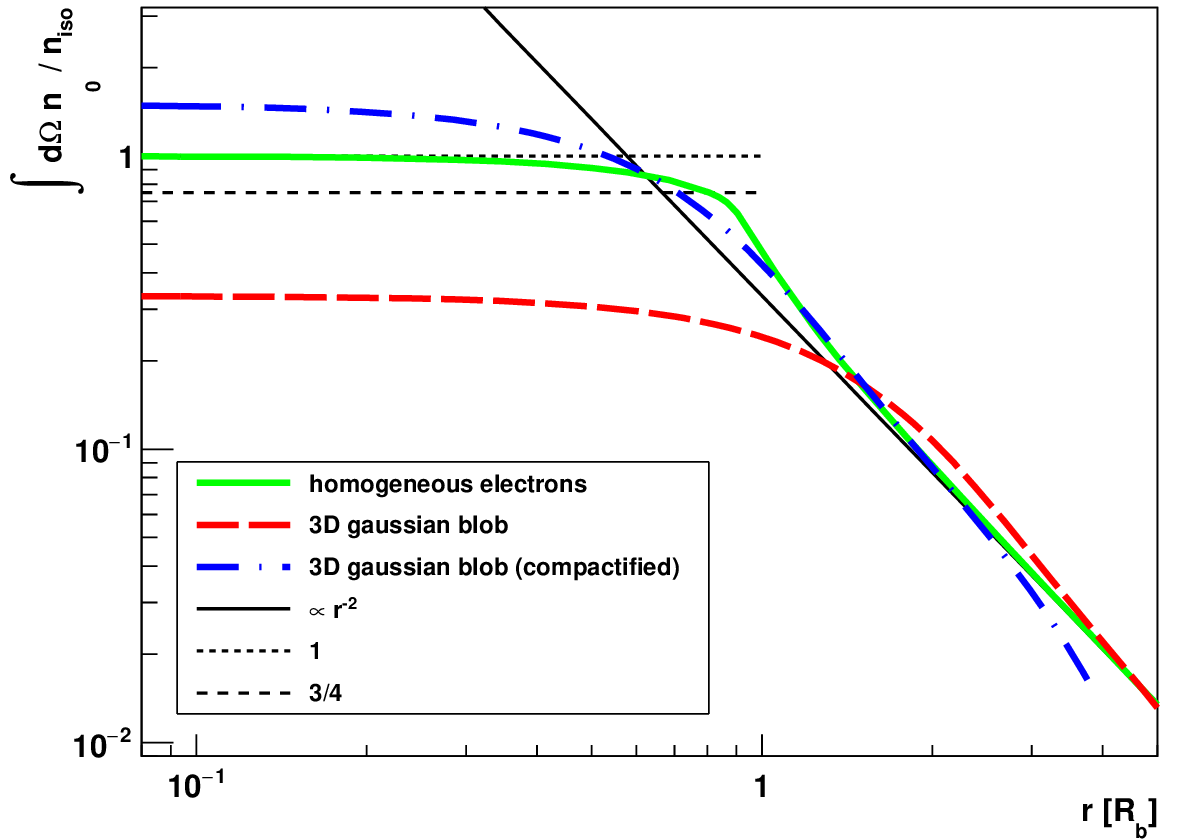}
  \caption{Total (integrated over all the directions) density of the radiation field as a function of the distance from the blob centre (normalized to the value at the centre of the blob for homogeneous blob, $n_{\rm iso}$) - thick lines (green solid for homogeneous electrons, red long-dashed for 3D Gaussian distribution of electrons, blue dot-dashed for 3D Gaussian compactified by factor $\sqrt{0.222}$ in radius. 
  Thin black lines show typical simplifications of density for homogeneous electrons:  value at the centre of the blob (dotted line), average density (dashed), point-like source (solid)
.}\label{fig:n_r}
\end{figure}

The total density of the radiation field we can obtain by integrating Eq.~\ref{eq:n0} over all the directions:
\begin{equation}
    \hat{n}_0(r, \epsilon_0) = \int_{-1}^1d\mu_s \int_0^{2\pi} d\phi_s n_0(r, \Omega, \epsilon_0)
\end{equation}
The result is compared with a few commonly used simplifications in Fig.~\ref{fig:n_r}.
This results has been cross checked to be consistent with the eq.~4 of \citet{1979A&A....76..306G}. 
The density of the radiation field can be considered nearly constant up to $r\lesssim0.5$, and for $r\gtrsim1.5$ it can be approximated by a point-like source.
However, for $r\approx1$, both approximations result in values about 30\% different from the full calculations.
It is worth noticing that a homogeneous blob will have a distribution of electrons $dN_e / dr \sim r^2$, resulting in the dominant effect of the highest values of $r$.

\begin{figure}
  \includegraphics[width=0.49\textwidth]{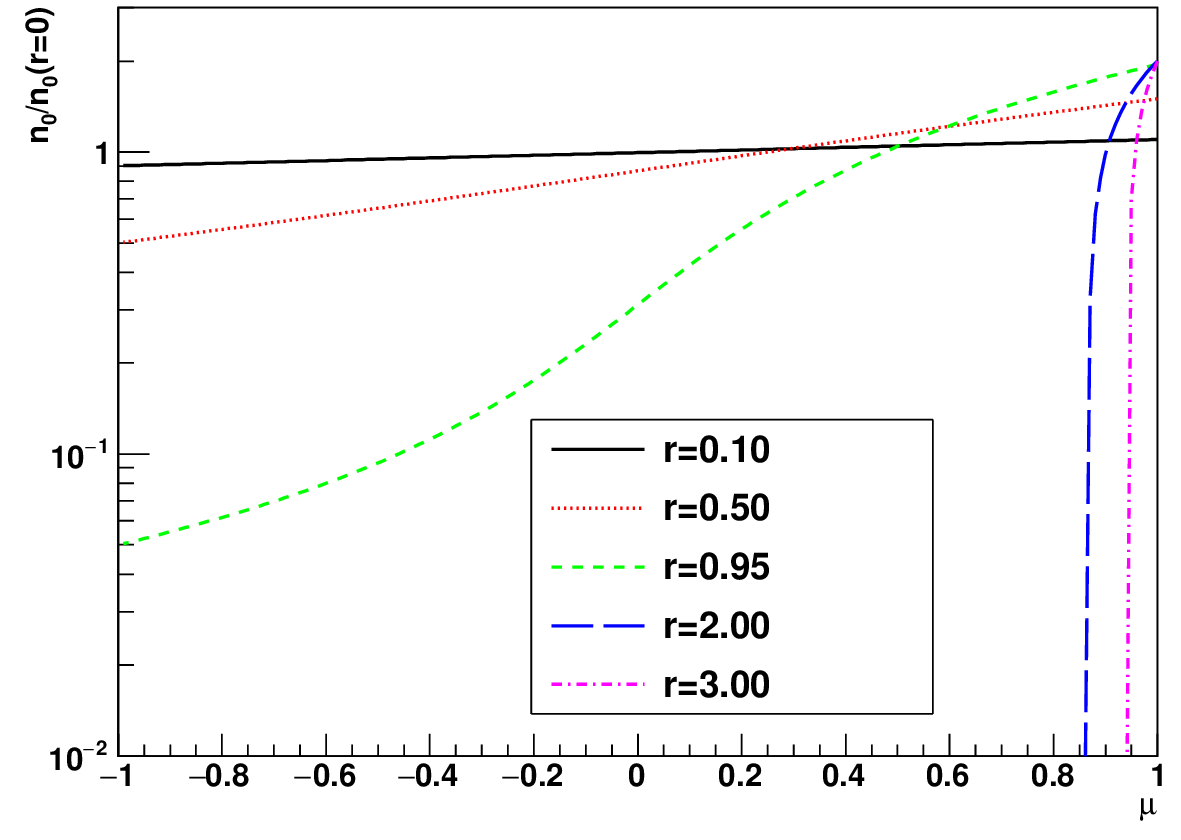}
  \caption{Distribution of the cosine of the angle (see Eq.~\ref{eq:n0}) to the radial direction ($\mu=\cos\theta$) of the radiation field normalized to such distribution at the centre of the blob ($n_0(r=0)$). 
  The angle $\theta$ can be interpreted as $\theta_s$ in Fig.~\ref{fig::sketch} for assumption of $\theta_o=0$.
  Different line colours and style correspond to different distance $r$ from the centre of the blob (see legend). 
}\label{fig:n_mu}
\end{figure}

In Fig.~\ref{fig:n_mu} we present the angular distribution of the radiation field seen at different distances from the centre of the blob. 

Moving away from the blob's centre, the distribution becomes increasingly anisotropic. 
Outside the blob, the radiation is observed only from a limited solid angle.

\subsection{SSC emission}
In order to compute the SSC emission we use the general formula of IC scattering of isotropic electrons on a directional radiation field (see  \citealp{aa81} and \citealp{mo05}).
We integrate such an emission over the whole anisotropic radiation field derived with Eq.~\ref{eq:n0}

\begin{figure}
    \centering
    \includegraphics[width=0.49\textwidth]{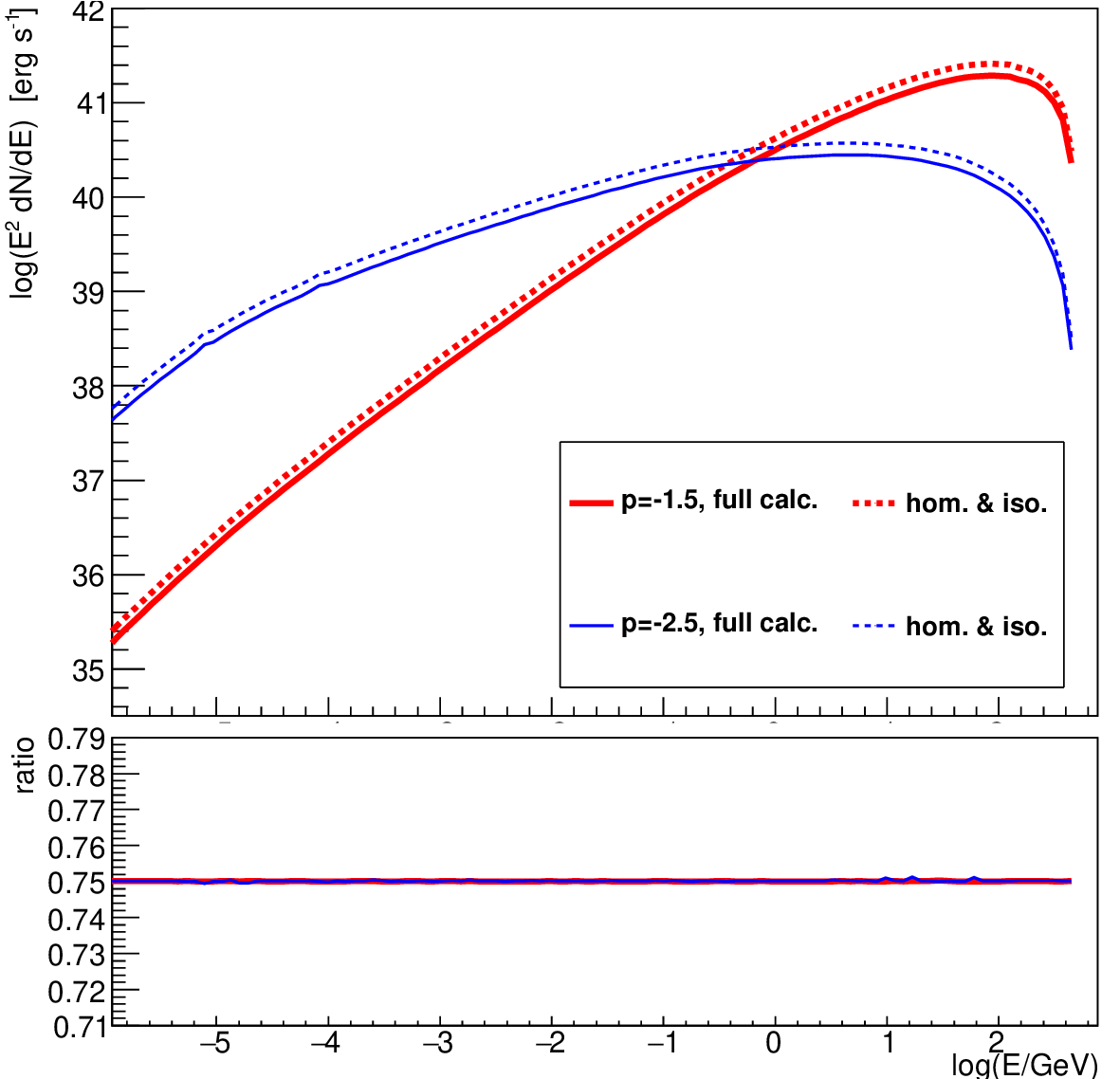}
    \caption{SED obtained in the SSC process for the full anisotropic inhomogeneous radiation field calculation (solid lines) and for a homogeneous and isotropic simplification (short dashed lines). Electron spectra spread between $\gamma=10^3$ and $10^6$ with index of $-1.5$ (red, thick) or $-2.5$ (blue, thin), $B=0.1$\,G, $R_b=10^{14}$\,cm. Electron Energy Distribution (EED) is normalized to the total power of $10^{44}$\,erg. 
    Bottom panel shows the ratio of the full calculations to the simplified case of isotropic and homogeneous radiation field} 
    \label{fig:ssc_sed}
\end{figure}

The total emission integrated over the whole blob is:
\begin{eqnarray}
  \frac{dN_\epsilon}{d\epsilon}&=
  \frac{3}{16\pi}c\sigma_T R_b^4
  \int_{-1}^1 2\pi d\mu_o
  \int_{-1}^1 d\mu_s\int_0^{2\pi} d\phi_s
  \int_0^1 dr 4\pi r^2 x \times \nonumber   \\ 
  & \int_{\ln \gamma_1}^{\ln \gamma_2} d(\ln\gamma) \gamma n_e(\gamma) 
   \int_{\ln \epsilon_{0,m}} d(\ln\epsilon_0) \epsilon_0 \frac{ \dot n_{0}}{4 \pi c \epsilon_0\gamma^2}
  f(\epsilon, \epsilon_0, \gamma, \theta_s) \label{eq:dne_sph}
\end{eqnarray}
where 
$ f(\epsilon, \epsilon_0, \gamma, \theta_s)$ is Eq. A3 in \citealp{mo05} and $x$ (dependent on $r$, $\mu_o$, $\mu_s$ and $\phi_s$) is computed according to Eq.~\ref{eq:x}.
For speed reasons, the two internal integrals (over the logarithms of the electron and soft photon energies) are tabularized as a function of $\log \epsilon$ and $\mu_o^4$. 
It is worth noting that the assumption of an isotropic and homogeneous synchrotron field in this formula would be equivalent to $x\equiv 1$ (calculations with this simplifications are shown with dotted lines in Fig.~\ref{fig:ssc_sed}). 

Sample calculations of the SSC emission and their comparison with the calculations in the case of  isotropic and homogeneous radiation field assumption are shown in Fig.~\ref{fig:ssc_sed}.
The full calculations, following Eq.~\ref{eq:dne_sph}, taking into account the inhomogenuity and the anisotropy of the radiation field, do not change the shape of the obtained SSC spectrum with respect to simplified calculations with the average, isotropic radiation field. 
This is not surprising, since the electron distribution is still assumed to be isotropic in the blob.
Hence, even with an anisotropic radiation field, for the total integrated emission, the distribution of the angles between the electrons and soft radiation is still isotropic.
Therefore, it is sufficient to apply the scaling factor of $3/4$, originating from the averaging of the soft radiation field density over the volume of the blob, to properly reproduce the SSC emission. 

\begin{figure}
    \centering
    \includegraphics[width=0.49\textwidth]{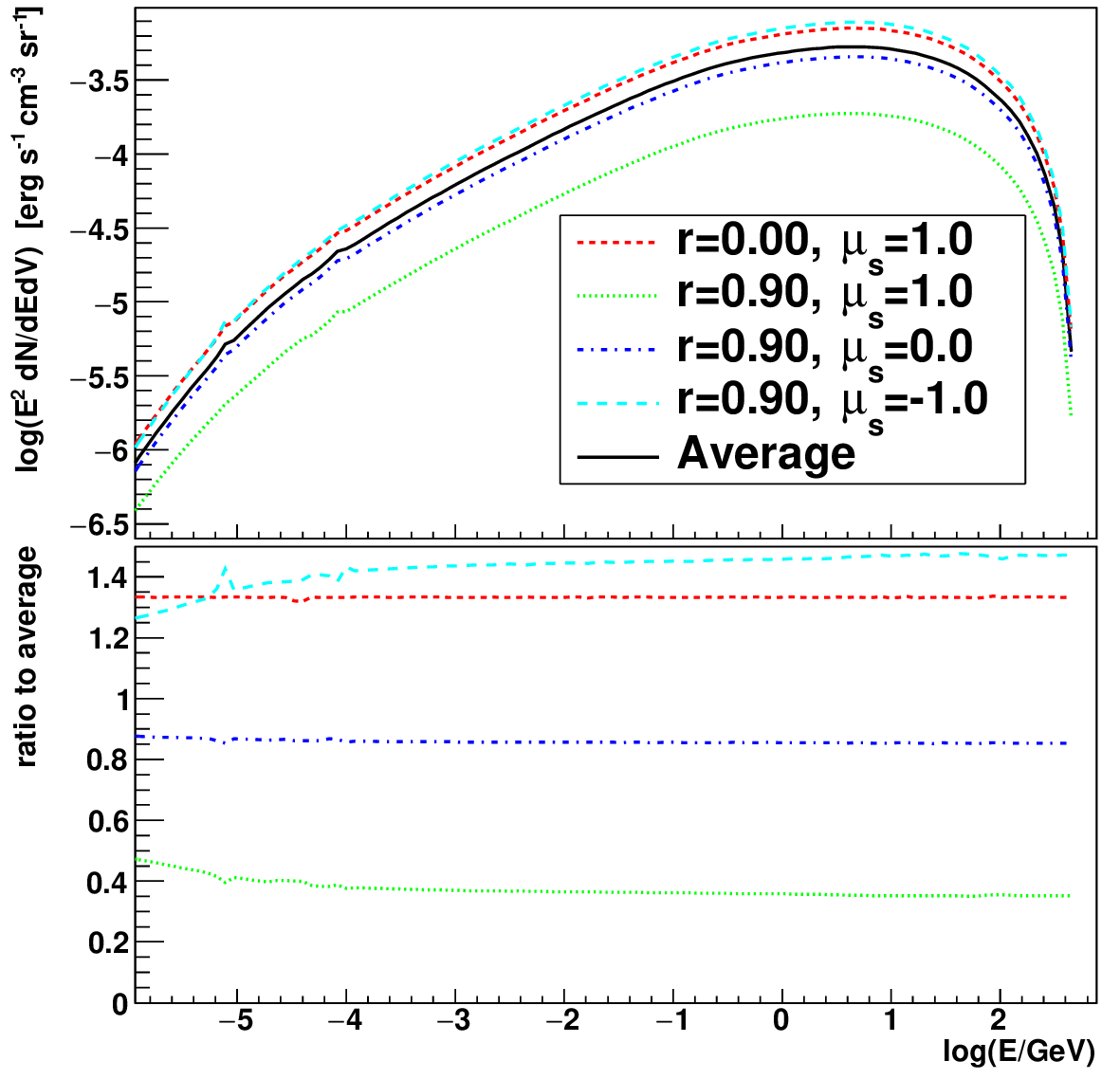}
    \caption{Top panel: density of the SSC emission from different places in the blob emitted at different angles (see the colours and line styles in the legend). Thick gray line shows the average emission integrated over the whole blob and directions and the  bottom panel shows the ratio to this average emission. Electron spectra spread between $\gamma=10^3$ and $10^6$ with an index $-2.5$, $B=0.1$\,G, $R_b=10^{14}$\,cm. EED is normalized to the total power of $10^{44}$\,erg.}
    \label{fig:ssc_angles}
\end{figure}

In Fig.~\ref{fig:ssc_angles} we present the SSC emission from various places of the blob emitted at different angles. 
Due to different radiation field density and angular distribution, the (directed) IC emission from various locations within the emission region can vary significantly, including a minor energy dependence. 

\subsection{Absorption in synchrotron radiation}\label{sec:abs}

A common simplification (see e.g. \citealp{fi08,2022A&A...660A..18N}) for a rough estimation of the effect of absorption of gamma-ray emission in the synchrotron radiation is to assume that the radiation field is homogeneous and only affects the high-energy radiation within the volume of the emission region.
We first compute a simplified optical depth, corresponding to an absorption through a constant layer of radiation field, $\tau^*(\epsilon)$:
\begin{equation}
    \tau^*(\epsilon) = 2R_b\int d\epsilon_0 \sigma_{\gamma\gamma}  \frac{3\dot n_0 (\epsilon_0) R_b}{4 c}\label{eq:tausimple},
\end{equation}
where $\sigma_{\gamma\gamma}=\sigma_{\gamma\gamma}(\epsilon, \epsilon_0, \theta_s)$ is the energy-dependent pair production cross-section. 
The 3/4 factor comes from averaging of the radiation field over the blob (see the previous section). 
For the purpose of calculating the optical depth with Eq.~\ref{eq:tausimple} we assumed that the absorption occurs throughout the diameter of the blob and that the photons interact at the $\theta_s=\pi/2$ angle. 
Notably \citet{1967PhRv..155.1404G} derived it also for an isotropic distribution of photons (see also \citealp{fi08}).
In the case of homogeneously emitting and absorbing region (a slab of material), the total attenuation can be computed as (see e.g. \citealp{fi08}):
\begin{equation}
    A^* = (1-e^{-\tau^*})/\tau^*. \label{eq:tauhom}
\end{equation}

\citet{1979A&A....76..306G} derived (however in the context of synchrotron-self-absorption) the absorption of a homogeneously emitting and absorbing spherical region. 
The attenuation factor in this case can be computed as (see \citealp{2009ApJ...692...32D}):
\begin{equation}
    A_{\rm sph}=\frac{3}{\tau^*}\left(
    \frac{1}{2}+
    \frac{e^{-\tau^*}}{\tau^*} - 
    \left(\frac{1-e^{-\tau^*}}{{\tau^*}^2}\right)
    \right)\label{eq:a_sph}
\end{equation}

If we take into account the inhomogenuity and the anisotropy of the radiation field, the optical depth for photons emitted at particular location inside the emission region and in particular direction is a function of not only the photon energy, but also its emission location ($r$) and direction ($\mu_o$), see Fig.~\ref{fig::sketch}.
\begin{eqnarray}
\tau(\epsilon, r, \mu_o)&=&
\int_0^\infty R_b dl \int_{-1}^1 d\mu_s \int_0^{2\pi} d\phi_s \int d(\ln\epsilon_0) \sigma_{\gamma\gamma} \nonumber \\
&\times& (1-\mu_s) n_0(\Omega(r',\mu'_0, l), \epsilon_0, r'(r, \mu_o, l))\epsilon_0
\end{eqnarray}
where $l$ (in units of $R_b$ is the integration along the path of the gamma ray, with the new position $r'=\sqrt{r^2+l^2+2rl\mu_o}$
and observation angle $\mu'_o=(l^2+r'^2-r^2)/(2lr')$.
In this case $\sigma_{\gamma\gamma}$ also depends on the scattering angle $\theta_s$. 

\begin{figure}
    \centering
    \includegraphics[width=0.49\textwidth]{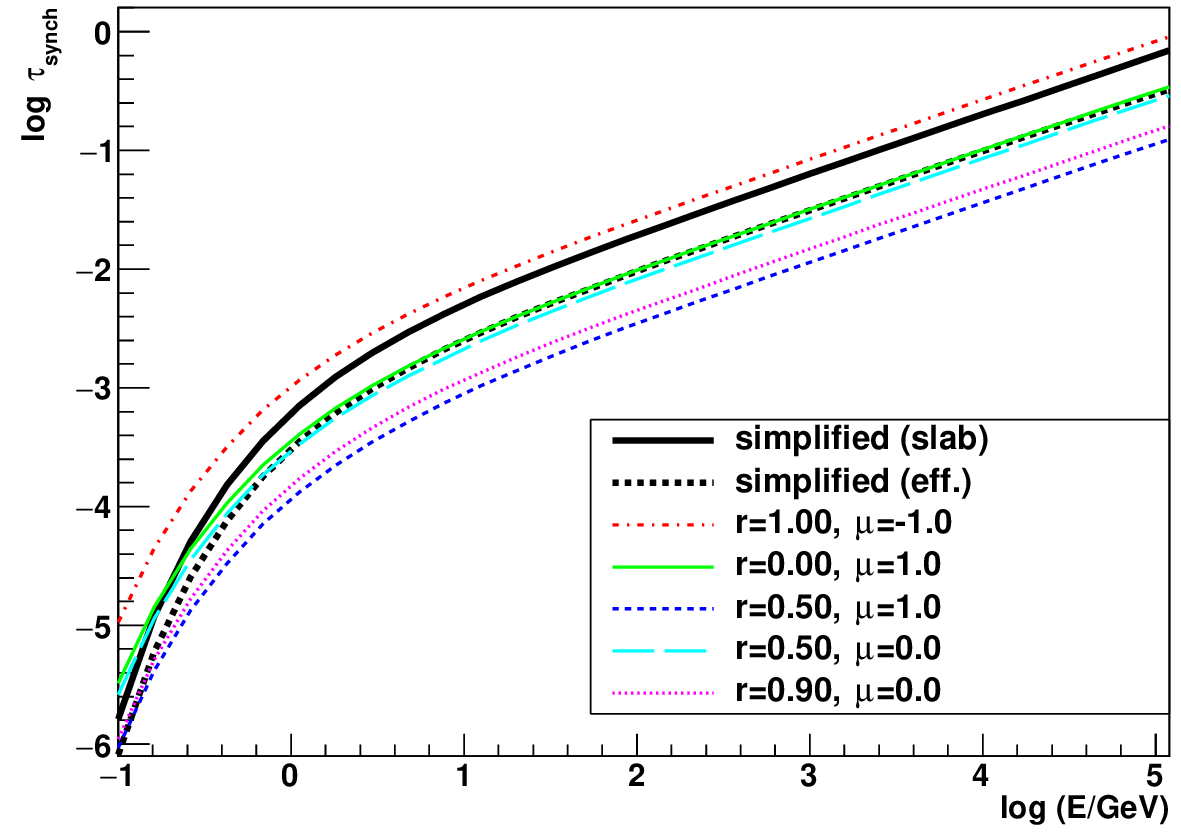}
    \caption{Absorption of gamma rays as a function of the production place $r$ and the cosine of polar angle of the photon, $\mu$ (see legend). For comparison in thick solid line simplified calculations, $\tau^*$, according to Eq.~\ref{eq:tausimple}, and thick dotted line the effective optical depth, $-\ln A^*$, corresponding to Eq.~\ref{eq:tauhom}.
    Electron spectra spread between $\gamma=10^3$ and $10^6$ with photon index of $-2$, $B=0.1$\,G, $R_b=10^{14}$\,cm. EED is normalized to the total power of $10^{44}$\,erg. }
    \label{fig:abs}
\end{figure}

In Fig.~\ref{fig:abs} we show example calculations of such optical depths for a few distances and directions and compare it to simplified calculations using Eq.~\ref{eq:tausimple}.
We neglect the effect of the IC emission by the secondary $e^+e^-$ pairs produced in the absorption.
Depending on the location of the gamma ray, the resulting optical depth can differ by about an order of magnitude. 
The bend in the absorption below 1~GeV is corresponding to the upper edge of the synchrotron emission produced by electrons. 
The geometrical dependence of the $e^+e^-$ pair production cross section on the interaction angle causes changes in the bend shape for different locations of the gamma ray.

The total observed emission will be the integral over the whole region and observation angles of the SSC photons from that particular emission location and at that particular angle times the corresponding absorption. 
Thus, in order to evaluate the total attenuation effect, we need to average over the emission produced within the blob:

\begin{equation}
    A_{\rm tot}=\frac{\int_0^1 dr r^2 \int_{-1}^1 d\mu_o F(r,\mu_o) e^{-\tau(\epsilon, r, \mu_o)}} 
    {\int_0^1 dr r^2 \int_{-1}^1 d\mu_o F(r,\mu_o)},\label{eq:a_tot}
\end{equation}
where $F(r,\mu_o)$ is the gamma-ray flux emission at the distance $r$ and at the angle $\arccos\mu_o$ to the radial direction.  
We note that contrary to the SSC emission and $\tau^*$, the non-linear terms in $A^*$ do not allow for a simple scaling with the density of the radiation field or the normalization of the electron distribution. 
Due to the dependence on $F$, the absorption is expected to affect differently gamma rays produced in different processes. 
If the emission is produced on a radiation field external to the blob (in the External Compton scenario) $F(r,\mu_o) \equiv 1$.

\begin{figure}
    \centering
    \includegraphics[width=0.49\textwidth]{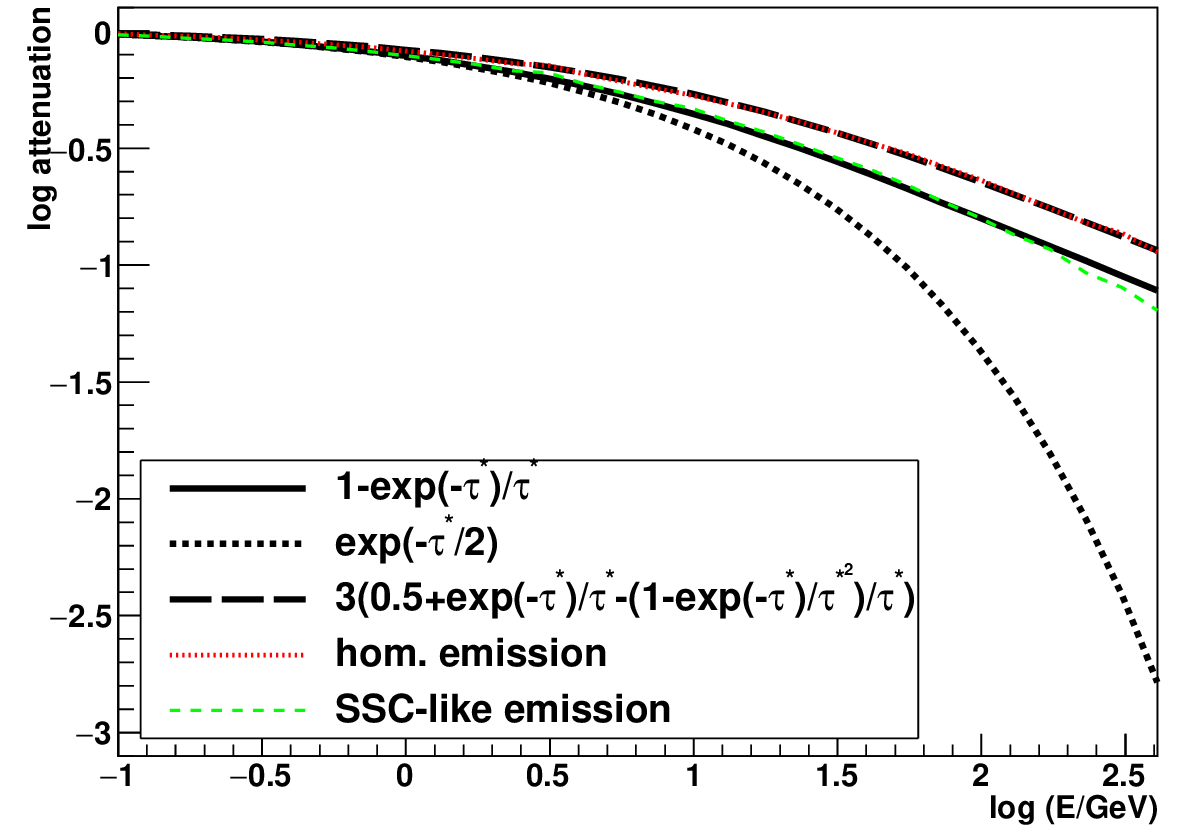}
    \caption{Thin lines: the average absorption factor in synchrotron radiation at the assumption of homogeneous emission of radiation (dotted red line) and taking into account the distribution of the emission (green dashed lines). 
    The electron spectrum spreads from $10^3$ to $10^6$ with an index of $-2$ and is normalized to $10^{45}$~erg. The magnetic field is set to 1~G. 
    For comparison the simplified calculations, $\exp(-\tau^*/2)$, see Eq.~\ref{eq:tausimple} are shown in thick solid line, the effective optical depth, $A^*$, corresponding to Eq.~\ref{eq:tauhom} is shown with the thick dotted line and Eq.~\ref{eq:a_sph} with the thick long dashed line.}
    \label{fig:att_comp}
\end{figure}

In Fig.~\ref{fig:att_comp} we present example comparison of the attenuation of gamma rays in the case of simplified calculations (homogeneous radiation field) and from full calculations of anisotropic and inhomogeneous radiation field. 
We compare it with the optical depth equal to half of Eq.~\ref{eq:tausimple} (while $\tau^*$ represents the optical depth going through the whole emission region, this happens only for the photons produced at the edge directly in the direction of the centre).
$\tau^*/2$ gives the optical depth of a photon produced in the centre, which is more representative when comparing with a homogeneously emitting region (considered in Eq.~\ref{eq:tauhom} and Eq.~\ref{eq:a_sph}). 

The numerical calculations in the case of homogeneously emitting and absorbing region are well reproducing the analytical calculation of Eq.~\ref{eq:a_sph} (compare red dotted and black long-dashed lines in Fig.~\ref{fig:att_comp}).
Inclusion of the effect of non-homogeneous production of gamma-ray radiation by homogeneously distributed electrons increases the absorption. 
This is expected because the production of gamma rays is enhanced either in the central parts of the emission region or in the outer parts for electrons pointing towards the centre of the emission region. 
In both of those cases, the absorption of the gamma rays is stronger than for outwards-pointing electrons.
In principle, the difference in the absorption could potentially be used to discriminate the source of the target radiation field (in particular distinguishing between the SSC and EC scenarios). 
However, in practice the difference is too small, and the IC spectrum will also be modified with the second component in the case when EC process accompanies the SSC emission. 
In the case of EC process the effect of anisotropy of the external radiation field is typically even more pronounced than for internal radiation field in SSC. 

Intriguingly, the full calculations result in an effective attenuation of the emission quite similar to that obtained from Eq.~\ref{eq:tauhom}. 
This makes it feasible and relatively safe to use such a simplification in the numerical codes that require fast calculations (such as fitting procedures).

\section{Inhomogeneous emission regions}\label{sec:inhom}
As a first step of generalization of the emission region's geometry, we consider a region that is filled with inhomogeneous distribution of electrons, which however is spherically symmetric (i.e. depends only on the radius from the centre of the region). 

\subsection{Inverse Compton Scattering}\label{sec:ic_gaus}
As long as the distribution of the emitting electrons is spherically symmetric, the density of the synchrotron radiation field will also have the same kind of symmetry (see e.g. the discussion in section IIIb) of \citet{1979A&A....76..306G}).
The total integrated emission, as seen by a distant observer from the whole region, will be isotropic. 
While at a given position the radiation field seen by the electrons will be anisotropic, the electrons are assumed to be isotropised by the magnetic field. 
Therefore, the distribution of the scattering angle will be the same as for the isotropic radiation field. 
Thus, the resulting IC emission for differently distributed electrons will simply scale with a constant factor that represents the integral of the product of the density of scattering electrons with the density of the radiation field that they encounter.

The scaling factor (normalized to emission expected from a homogeneous distribution of electrons, but taking into account the inhomogenuity of the resulting radiation field as in Section~\ref{sec:hom}) can be computed as:
\begin{equation}
F_{ih}= 2 \int_0^\infty dr r^2  \hat{n}_e(r) \int_{-1}^1 d\mu \int_0^\infty dl \hat{n}_e\left(r_s(r, \mu, l)\right).    \label{eq:f_inhom}
\end{equation}

Where $\hat{n}_e$ is normalized such that $\int_0^\infty dr 4\pi r^2 \hat{n}_e(r)$ = $4\pi /3$. 
The case of a homogeneous distribution of electrons considered in Section~\ref{sec:hom} ($\hat{n}_e(r)$ distribution being 1 for $r<1$ and 0 for $r>1$), taking into account the resulting inhomogeneous and anisotropic radiation field, would correspond to $F_{ih}=1$.
On the other hand, if one additionally substitutes the last integral in Eq.~\ref{eq:f_inhom} by radius to reproduce the simplification of the homogeneous radiation field, $F_{ih}=4/3$, i.e. reproducing the inverse of the $0.75$ factor obtained in Section~\ref{sec:hom}.

As an example, we compute such an effect of the inhomogeneity of the electron distribution for a 3-dimensional Gaussian distribution, $\hat{n}_e(r) = (\sqrt{2/\pi}/3) e^{-r^2/2}$.
Such a distribution is expected, e.g. in the case of a random-walk-like diffusion. 
In general the diffusion can lead to electrons of different energies occupying spreading up to different distances.
However, the combined effect of cooling dominated by either synchrotron or IC in Thomson regime, and the Bohm-like diffusion cancel out making the radius independent on the energy of electrons (see eq.~14 in \citealp{2024ApJ...967...93Z}).
From the numerical integration, we obtain $F_{Gauss3D} = 0.222$. 
The IC scattering is much less efficient in this case than in the top-hat homogeneous scenario because the distribution has long tails. 
Only $20\%$ of the electrons are contained within $r<1$ and $48\%$ within $r<1.5$.

\subsection{Pair production absorption}
\begin{figure*}
    \centering
    \includegraphics[width=0.49\textwidth]{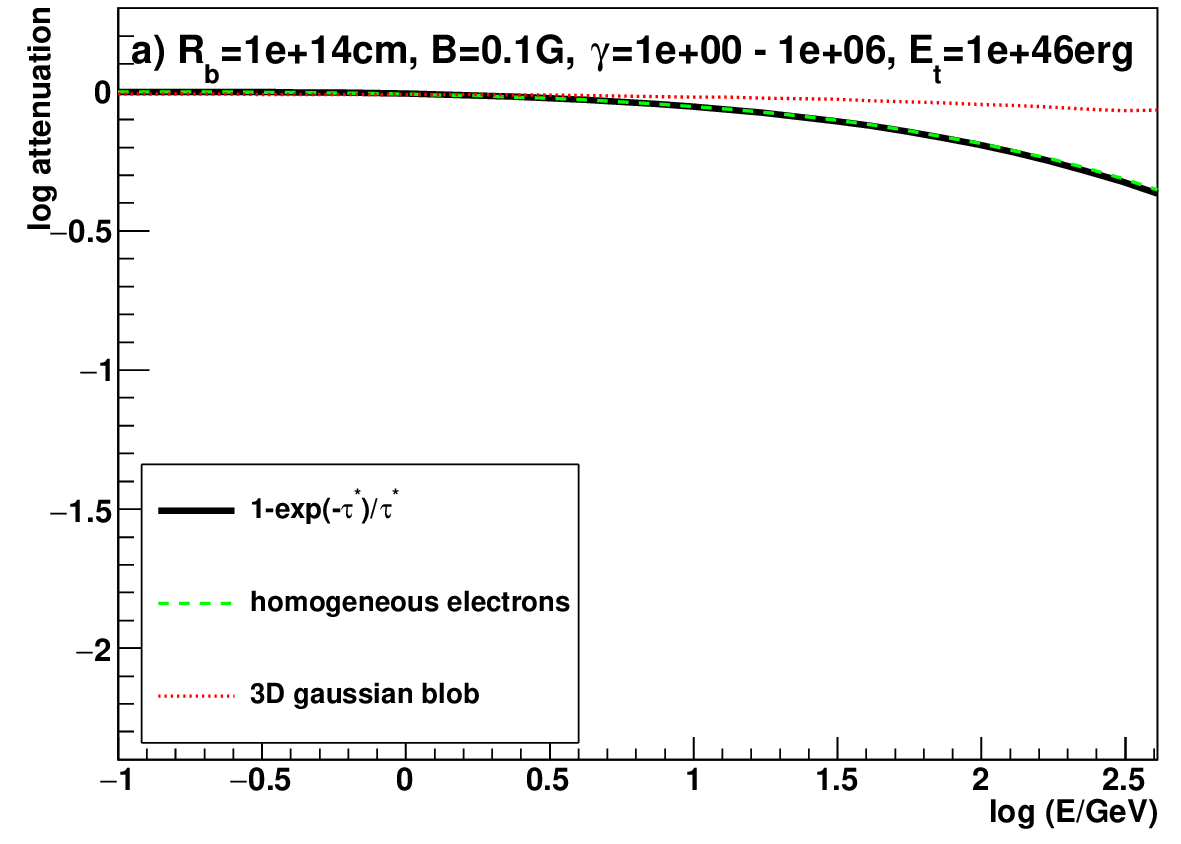}
    \includegraphics[width=0.49\textwidth]{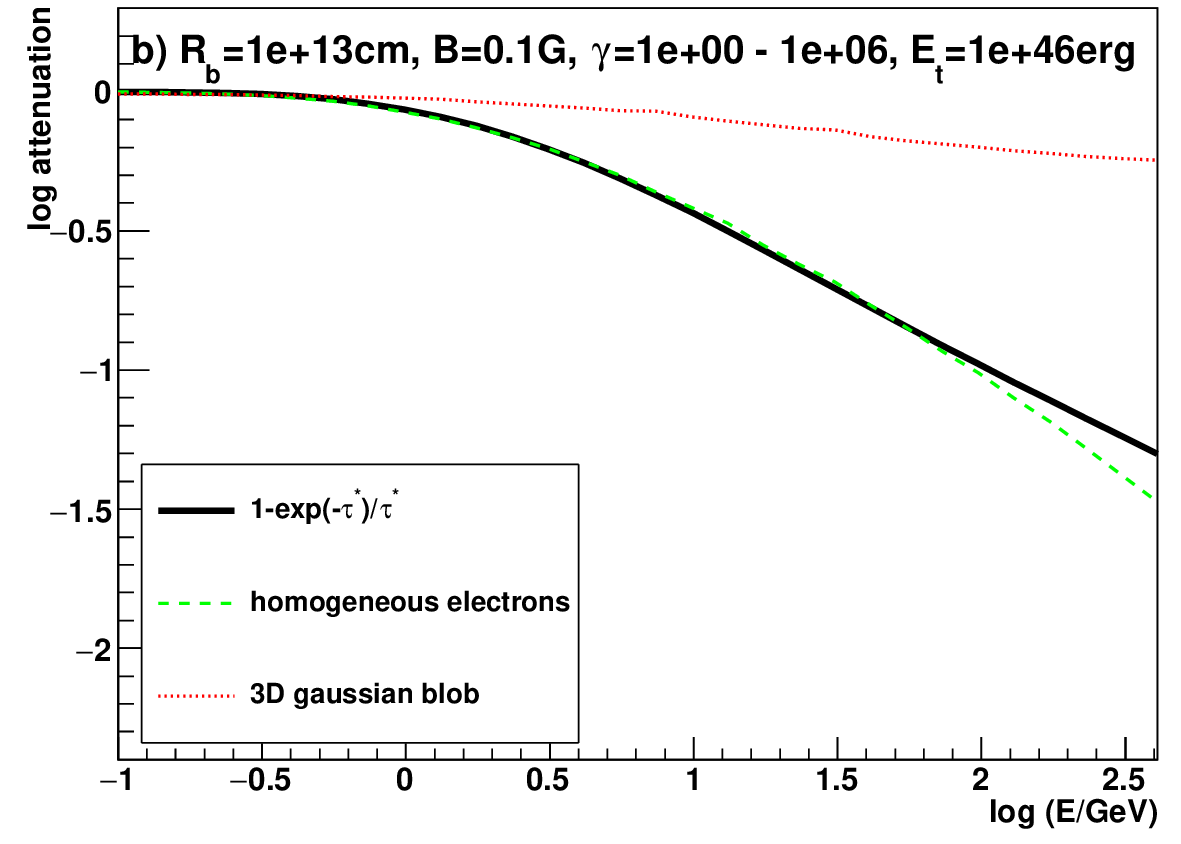}\\
    \includegraphics[width=0.49\textwidth]{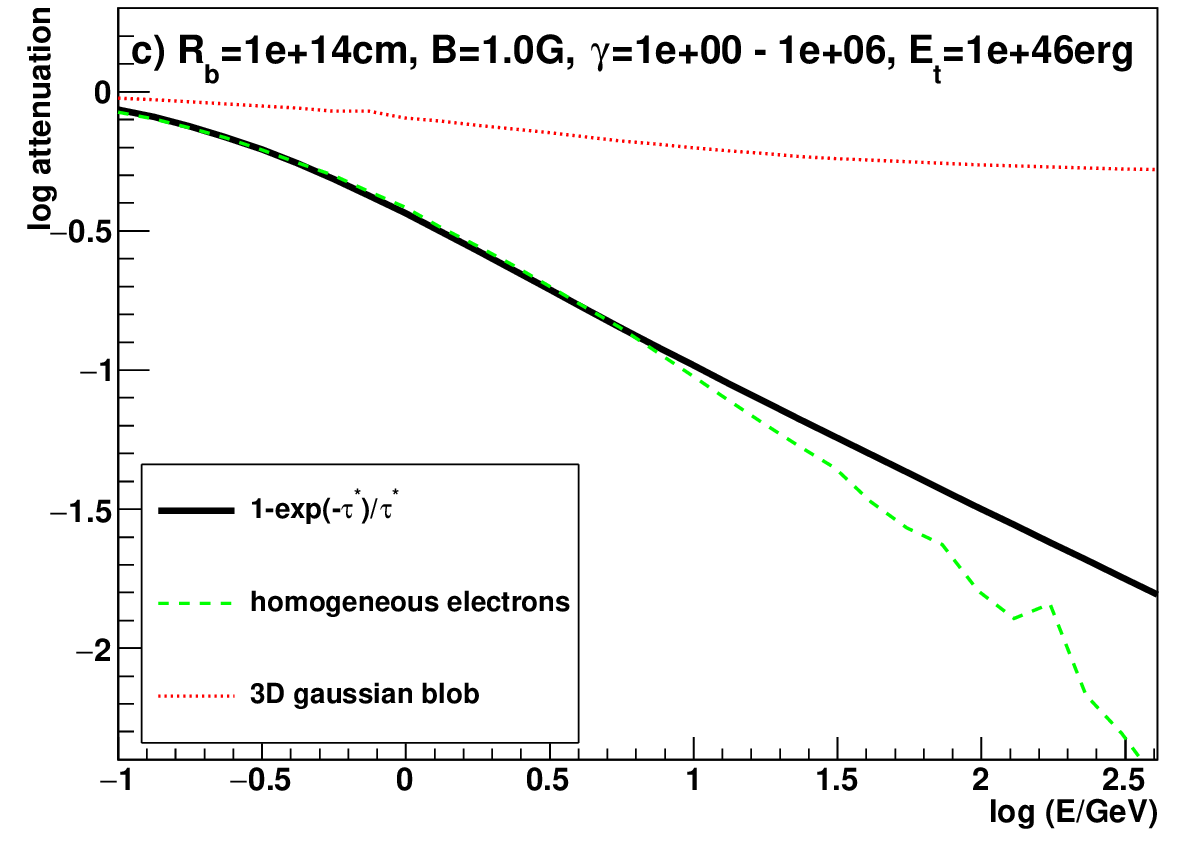}
    \includegraphics[width=0.49\textwidth]{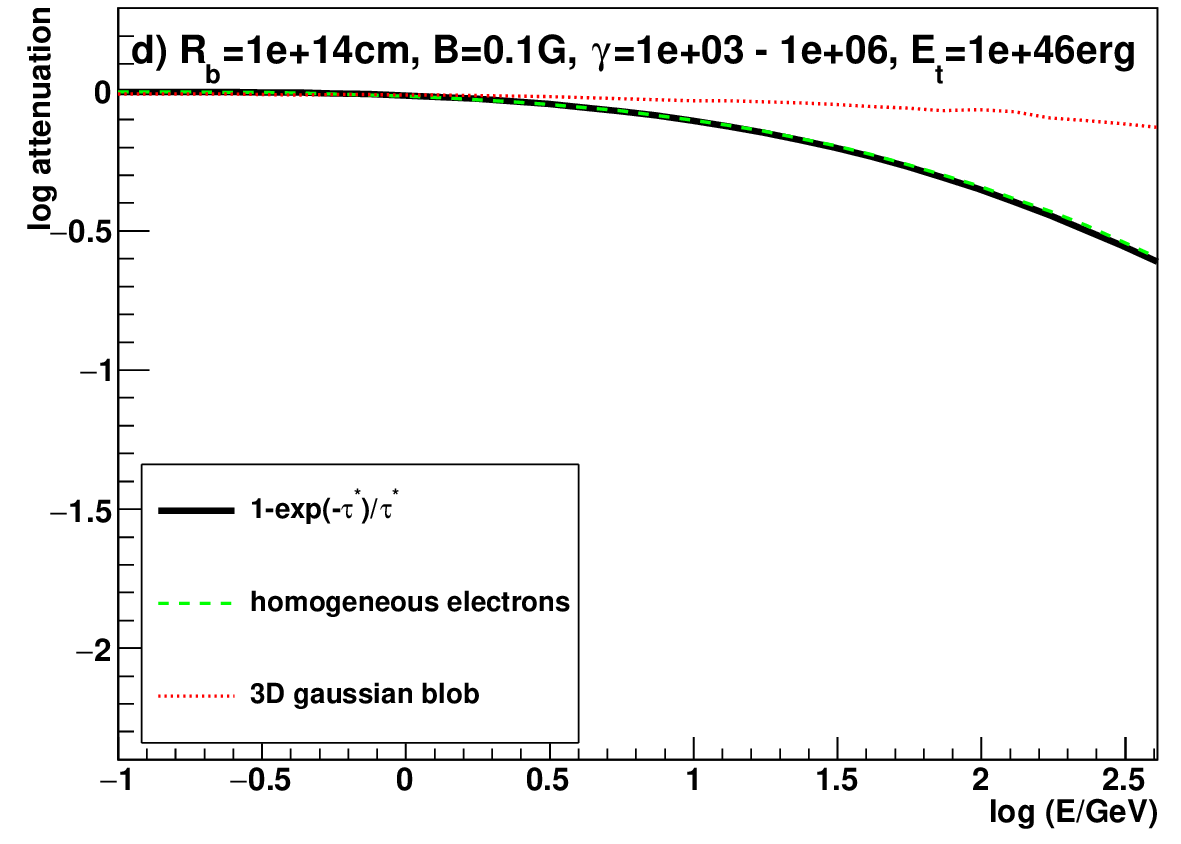}\\
    \includegraphics[width=0.49\textwidth]{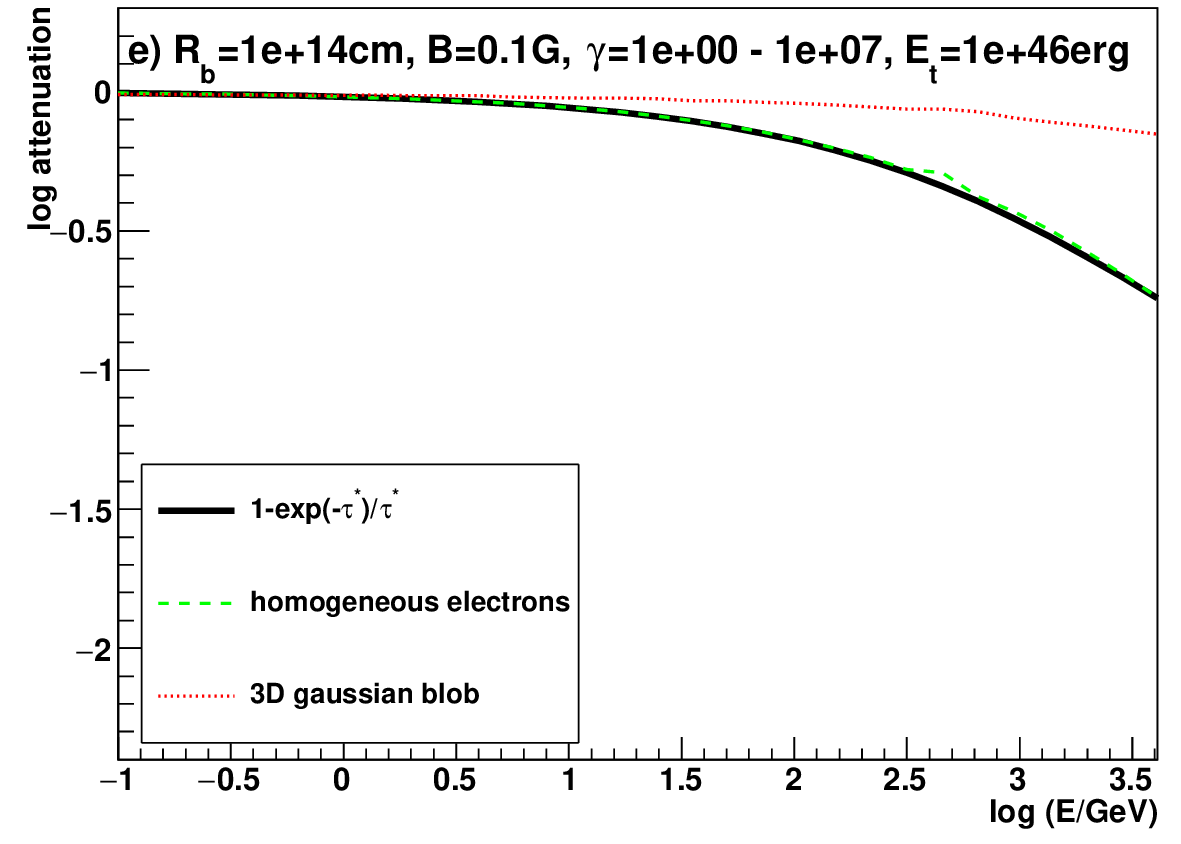}
    \includegraphics[width=0.49\textwidth]{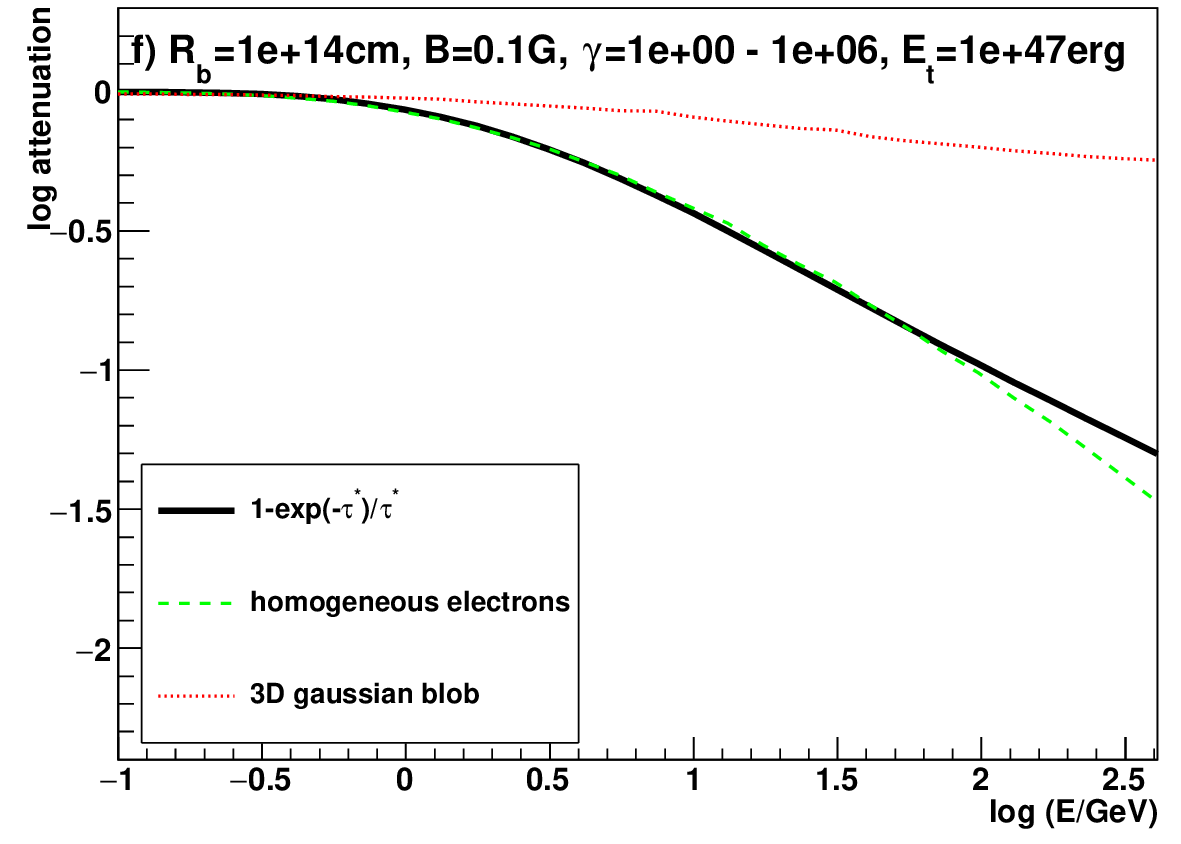}
    \caption{Comparison of the simplified average absorption from homogeneously emitting radiation (solid black line) with the full calculations of absorption of SSC radiation in the synchrotron radiation field for a homogeneously distributed electrons (flat-top distribution within the blob, green dashed lines) and electron density following a 3D Gaussian distribution (red dotted). The baseline scenario (a) is for radius of the blob $R_b=10^{14}$~cm, magnetic field $B=0.1$~G, and electron energy distribution between 1 and $10^6$ with a slope of $-2$ normalized to $10^{46}$~erg. Case (b) shows more compact blob with $R_b=10^{13}$~cm, case (c) stronger magnetic field ($B=1$~G), case (d) increased minimal Lorentz factor of the electrons to $10^3$, case (e)  increased maximum Lorentz factor to $10^7$ and case (f) a higher total normalization of the electrons $E_t=10^{47}$~erg.}
    \label{fig:abs_gaus}
\end{figure*}

The effect of absorption can be calculated in an analogous way to Eq.~\ref{eq:a_tot} (with a slight modification of the integration range in radius should now go to infinity). 
The calculations are more numerically complex (due to additional integral needed to derive $F(r,\mu)$), and to counteract this, we tabularize the geometrical factor scaling the radiation field as a function of the radius from the centre of the emission region and the polar angle.
Example calculations for different combinations of parameters are presented in Fig.~\ref{fig:abs_gaus}, modifying one by one the basic parameters of the SSC to increase the absorption (decreasing the radius, increasing the magnetic field, increasing the energy range of the spectrum, or its normalization).

In all investigated cases, the full calculations of the absorption for a homogeneous blob can be well approximated with simple absorption by homogeneously emitting medium (Eq.~\ref{eq:tauhom}).
The differences start to be visible for strong absorption (attenuation factor of at least 1 order of magnitude). 
Notably, in the case of such strong absorption, this process is normally accompanied by second-order inverse Compton scattering of secondary electrons and subsequent cascading. 

The absorption for the 3D Gaussian distribution of the electrons is weaker. 
This is to be expected, because both the radiation field and the electrons themselves are more distributed, resulting also in weaker IC emission in that case.

In order to take into account this effect, we compare the absorption also with a ``compactified'' 3D Gaussian emission region, i.e. with a region with a characteristic radius of $R_{b,c}=\sqrt{0.222} R_b$.
The radiation field inside the blob and hence also the IC emission scales as $\sim R_b^{-2}$.
Since, as we derived in Section~\ref{sec:ic_gaus}, for the same value of $R_b$, 3D Gaussian distribution of electrons results in $0.222$ times weaker IC flux, the IC emission from a homogeneous blob with radius of $R_b$ and 3D Gaussian blob compactified to characteristic radius of $R_{b,c}$ results in the same IC (as well as synchrotron) emission if all are normalized to the same total energy distribution of the electrons. 

\begin{figure*}
    \includegraphics[width=0.49\textwidth]{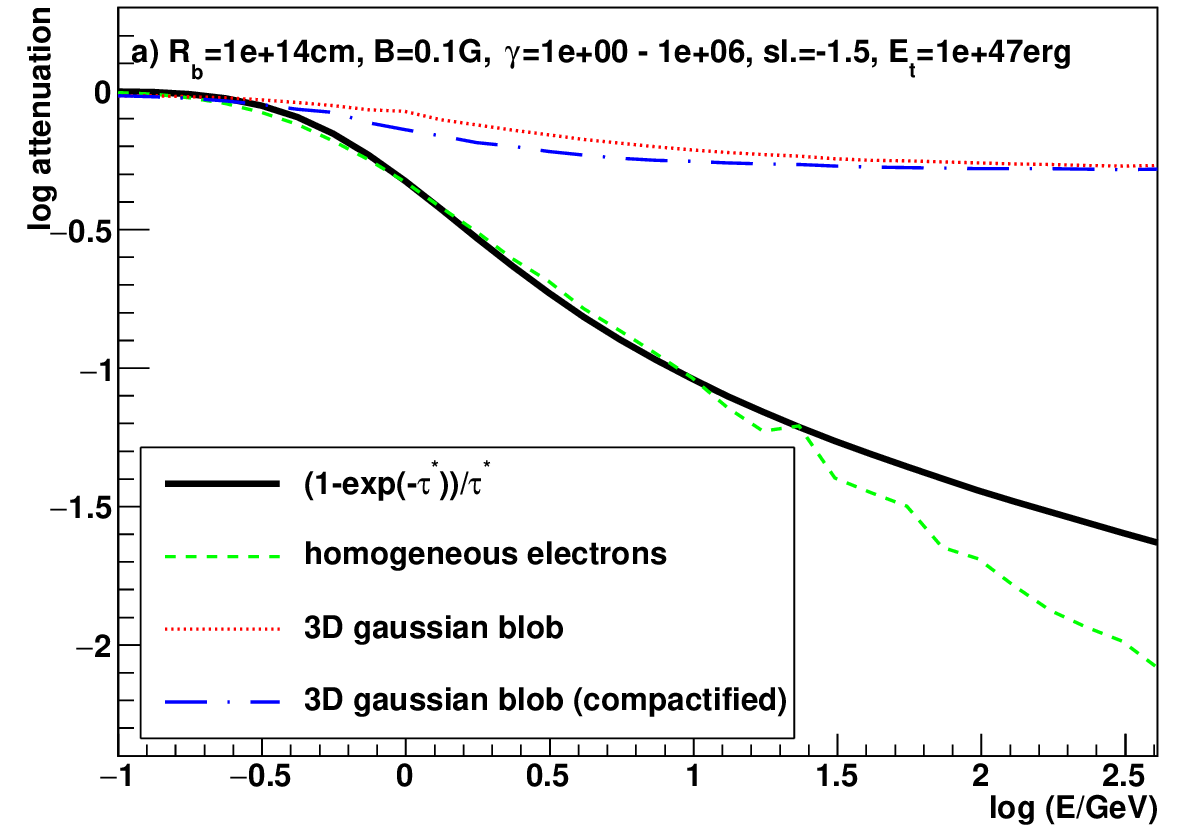}
    \includegraphics[width=0.49\textwidth]{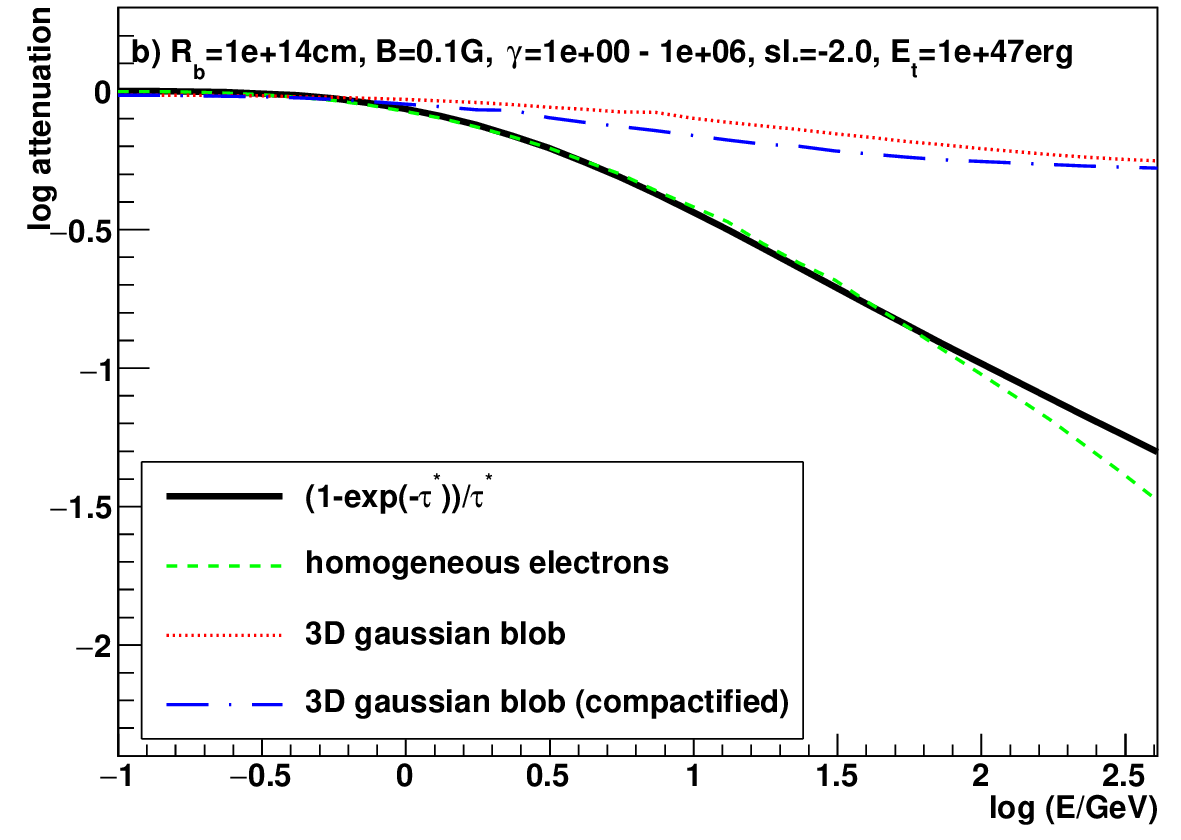}\\
    \includegraphics[width=0.49\textwidth]{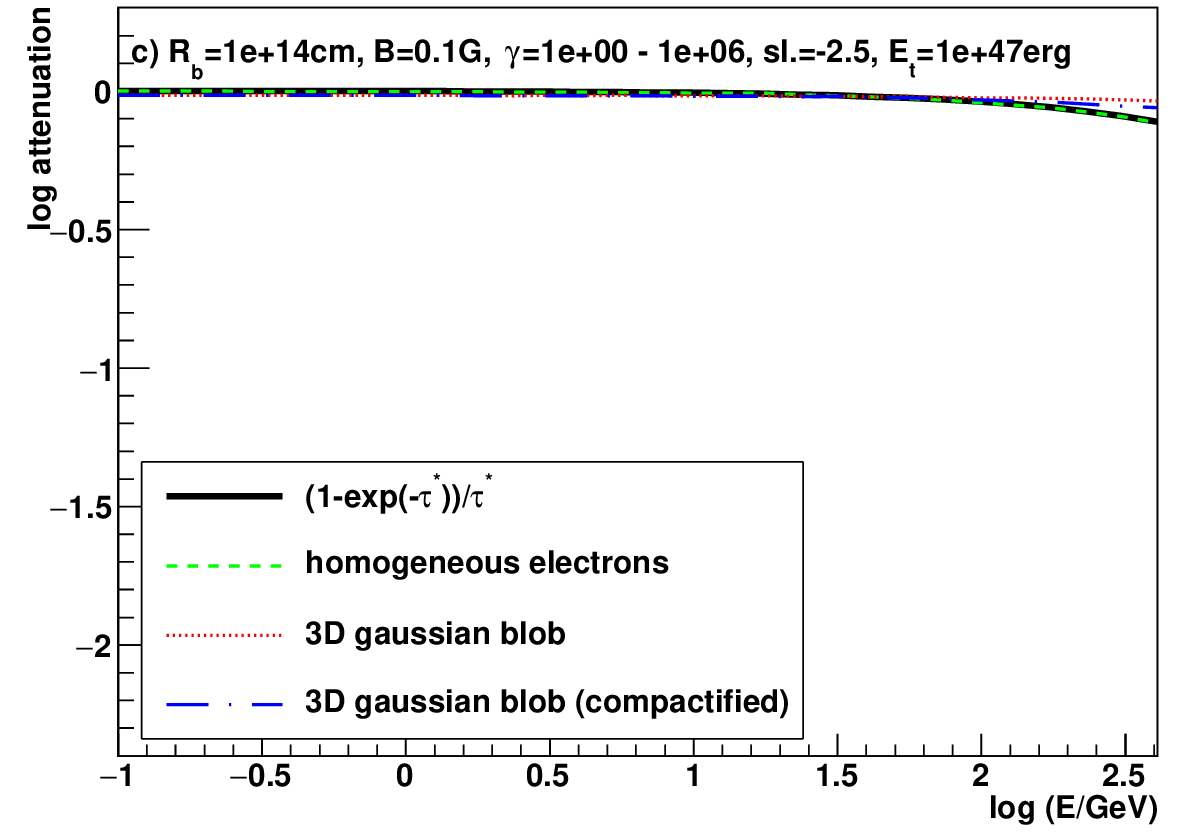}
    \caption{Comparison of the simplified average absorption from homogeneously emitting radiation (solid black line) with the full calculations of absorption of SSC radiation in the synchrotron radiation field for a homogeneously distributed electrons (flat-top distribution in the  blob, green dashed lines) and electron density following a 3D Gaussian distribution with the same radius (red dotted) or compactified to have the same IC emission (blue dot-dashed). 
    The radius of the homogeneous blob $R_b=10^{14}$~cm; magnetic field $B=0.1$~G, and electron energy distribution between 1 and $10^6$ with a slope of $-1.5$, $-2$ or $-2.5$  (top-left top-right and bottom panel respectively) is normalized to $10^{47}$~erg.}\label{fig:gaus_comp}
\end{figure*}

In Fig.~\ref{fig:gaus_comp} we show the comparison of the attenuation factor of such compactified 3D Gaussian, with the previous two scenarios, for different slopes of the electron distribution. 
The compactified blob naturally results in stronger absorption than the baseline scenario of using the same characteristic radius as the radius of the homogeneous blob.
However, despite the same synchrotron and IC emission in both cases, the absorption is still much weaker than that of the homogeneous flat-top distribution in the blob. 
This is likely due to the fact that a significant fraction of the IC emission in the 3D Gaussian blob is produced at its tails, where the absorption is already much weaker.

\section{Cylindrical emission region}

\begin{figure}
    \centering
    \includegraphics[width=0.49\textwidth]{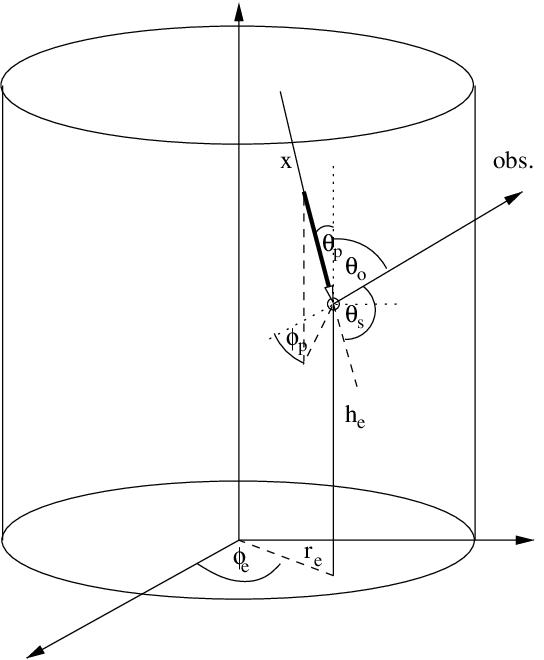}
    \caption{The geometry of cylindrical region. Scattering electron is marked with a circle and its position is determined by $r_e$, $h_e$ and $\phi_e$. 
    The thick arrow shows the radiation field incoming from the direction described by polar angle $\theta_p$ and azimuthal angle $\phi_p$.
    The observer is located in YZ plane at the angle $\theta_o$ to z axis. 
    The scattering angle is $\theta_s$.
    }
    \label{fig:cyl_geom}
\end{figure}

Next, we consider an opposite generalization of the emission region to the one discussed in Section~\ref{sec:inhom}.
Namely, now the distribution of the electrons does not have a spherical symmetry, but rather a cylindrical one, but for simplicity we assume that within this cylinder the density of the electrons is homogeneous. 
The assumed geometry of the cylindrical emission region is explained in Fig.~\ref{fig:cyl_geom}.

\subsection{Inverse Compton Scattering}
\begin{figure}
    \centering
    \includegraphics[width=0.49\textwidth]{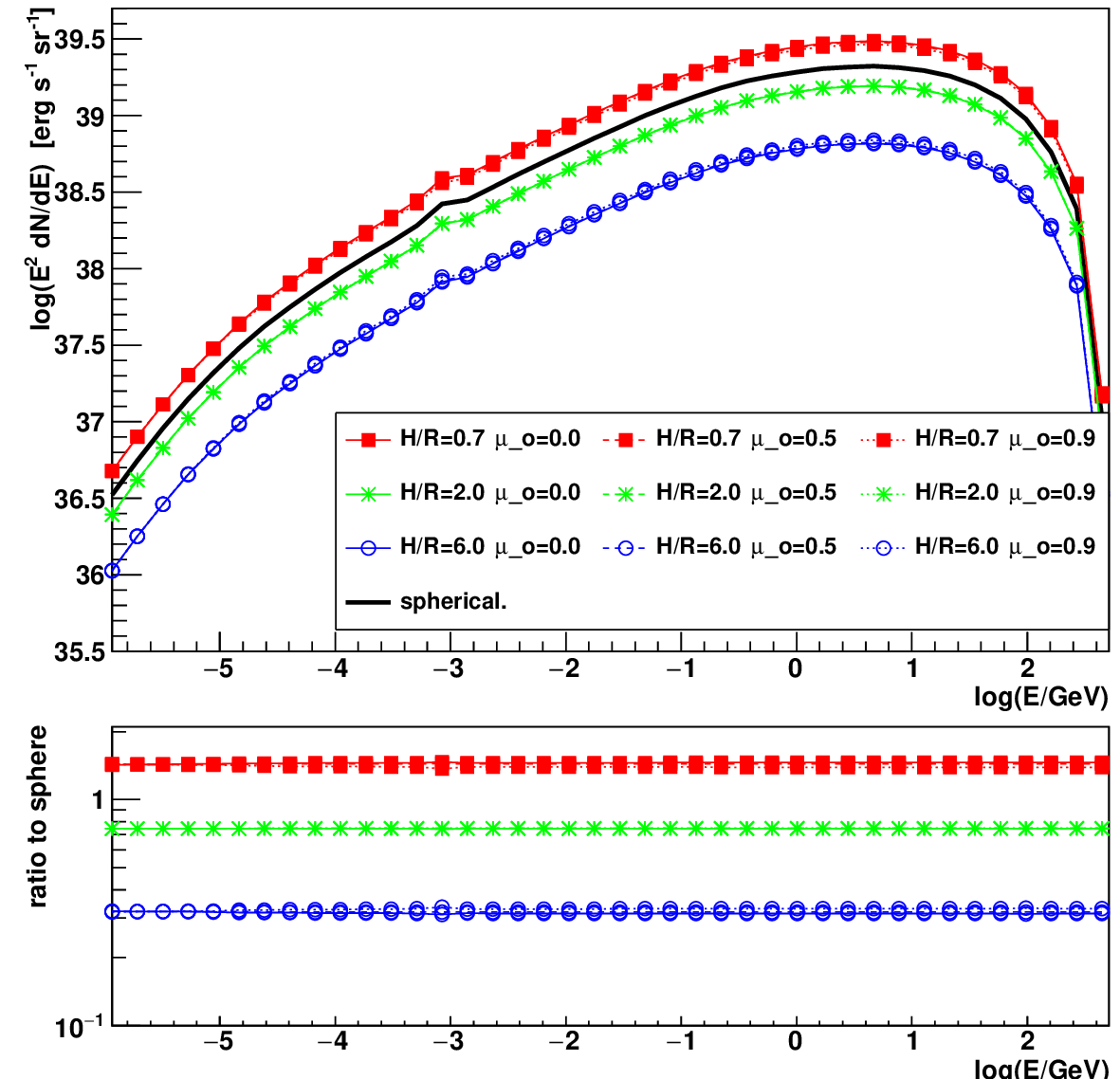}
    \caption{SED of SSC emission obtained from an emission region with a radius of $10^{14}$~cm filled with $B=0.1$~G magnetic field and electron energy distribution spreading from $\gamma=10^3$ to $10^6$. The height to radius ratio of the cylinder is set to 0.7 (red lines and full squares), 2 (green lines and stars) and 6 (blue lines and empty circles). 
    The cosine of the observation angle is set to 0 (solid lines), 0.5 (dashed) and 0.9 (dotted), however for a given H/R ratio, the three curves corresponding to different observation angels overlap within the numerical precision of the calculations and cannot be distinguished in the plot.
    Emission from a spherical blob is shown with thick black line. 
    In all the cases the total energy of electrons is normalized to $10^{44}$~erg. 
    The bottom panel shows the ratios of the spectra to the one from spherical blob.}
    \label{fig:sed_cyl}
\end{figure}

In order to calculate the equivalent of Eq.~\ref{eq:dne_sph} for a cylindrical blob, we need to integrate over the whole volume of the scattering electrons (height $h_e$, radius $r_e$ and azimuthal angle $\phi_e$) as well as over the direction of the incoming photons (cos of polar angle $\mu_p$ and azimuthal angle $\phi_p$). Due to a lack of evident spherical symmetry, we do not integrate over the observation angle $\theta_o$. 

\begin{eqnarray}
  \frac{dN_\epsilon}{d\epsilon d\Omega_o}&=
  \frac{3}{16\pi}c\sigma_T 
  \int_0^H dh_e
  \int_0^R r_e dr_e 
  \int_0^{2\pi} d\phi_e
  \int_0^{2\pi} d\phi_p
  \int_{-1}^1 d\mu_p x \nonumber   \\
  \times & \int_{\ln \gamma_1}^{\ln \gamma_2} d(\ln\gamma) \gamma n_e(\gamma) 
   \int_{\ln \epsilon_{0,m}} d(\ln\epsilon_0) \epsilon_0 
   \frac{ \dot n_{0} f(\epsilon, \epsilon_0, \gamma, \theta_s)}{4 \pi c \epsilon_0\gamma^2}
\end{eqnarray}
The cosine of the scattering angle, $\mu_s=\cos\theta_s$ can be computed as
\begin{equation}
    \mu_s = -\sin\theta_o\sin\theta_p\cos\phi_p - \mu_p\mu_o \label{eq:musc}
\end{equation}
Analogically to the spherical case, $x$ is the distance to the edge of the cylinder and it can be computed as a minimal distance from the top/bottom (for $\mu_p>0$ and $\mu_p<0$ respectively) and from the side:
\begin{eqnarray}
    x_t&=&(H-h_e)/\mu_p \label{eq:xt}\\
    x_b&=&-h_e/\mu_p \\
    x_s&=&\frac{\pm\sqrt{R^2-r_e^2\sin^2(\phi_p-\phi_e)}-r_e\cos(\phi_p - \phi_e)}{\sin\theta_p} \label{eq:xs}
\end{eqnarray}
For the position within the cylinder the solution with $+$ should be taken for $x_s$. 
In Fig.~\ref{fig:sed_cyl} we compare the SEDs obtained from cylinders with a different H/R observed at different angles with respect to their axis with the one obtained from a spherical blob.
As all emission regions are normalized to the same total energy of electrons, their synchrotron emission observed by a distant observers by definition will be the same. 
Nevertheless, the level of the IC emission depends on the H/R of the blob with the narrow cylinders having smaller emission than broad (flat) ones. 
This effect is the interplay of varying electron density and escape time of the synchrotron radiation from the blob, both affecting the emission in the opposite way. 
Namely a compression of the cylinder towards a flat disk will increase the electron density, that is relevant both in terms of the synchrotron emission as well as inverse Compton scattering. 
At the same time, radiation escape will be facilitated in the ``height'' dimension, partially counteracting the effect of the enhanced electron density.
On the other hand, in the case of an elongated cylinder, the electron density drops, but the escape of radiation from the region is still determined by the two horizontal ``radial'' dimensions. 
The flux enhancement factor compared to the homogeneous sphere with a radius of $R$ in the range of $H/2R$ in between $0.2$ and $20$ can be well described by a fit function:
\begin{equation}
    F_{cyl}(H, R)/F_{sph}(R)=0.75 (H/2R)^{-0.69-0.078\ln{H/2R}}.
\end{equation}
Since the density of the electrons in the emission region scales in this case as $1/H$, the emission scales as the electron energy density to an index of $\sim0.69$. 
This is a similar behaviour to an index of $2/3$ expected for the dependence of the flux on the electron energy density with total energy fixed, but varying radius of the sphere. 

Interestingly, contrary to naive expectations, the lack of spherical symmetry of the cylinder does not introduce any variation of the emission on the observation angle.
The numerical derivation of this effect for a simplified case is presented in the Appendix~\ref{sec:cylgeom}.

\subsection{Pair production absorption}

In order to compute the effect of the absorption for a cylindrical emission region we will take a similar approach like in Section~\ref{sec:abs}.
We first investigate the absorption of the radiation emitted from different parts of the emission region, and at different directions.
Due to cylindrical symmetry the optical depth $\tau$ will depend on five variables: energy $\epsilon$, the observation angle $\theta_o$, and on the location in the cylinder expressed with $r_e$, $\phi_e$ and $h_e$:
\begin{equation}
\tau=
  \int_0^\infty \!dl 
  \int_{-1}^1 \!d\mu_p \int_0^{2\pi} \!d\phi_p
   \frac{x' (1-\mu_s)}{4 \pi c}
\int \!d(\ln\epsilon_0) \epsilon_0\sigma_{\gamma\gamma} 
   \dot n_{0},
\end{equation}
where $x'$ is the length of the line within the cylinder along the line defined by angles $\phi_p'$ and $\theta_p'$. 
All these primed quantities are calculated after the gamma ray traverses the distance $l$ from its production location (described by $r_e$, $\phi_e$ and $h_e$), and can be computed as: 
\begin{eqnarray}
r_e'&=&\sqrt{r_e^2+2r_e l \sin\phi_e\sqrt{1-\mu_o^2}+l^2(1-\mu_o^2)}\label{eq:rep}\\
\tan\phi_e'&=&\frac{r_e\sin\phi_e+l\sqrt{1-\mu_o^2}}{r_e\cos\phi_e}\\
    h_e'&=&h_e+l\mu_o\label{eq:hep}
\end{eqnarray}
Using this system of coordinates the cosine of the interaction angle between the photons, $\mu_{sc}$ does not change with $l$ and can be computed directly with Eq.~\ref{eq:musc}.

Calculation of $x'$ is more complicated, in particular when the photon position is already outside of the cylinder. 
We calculate the crossing points with the surfaces containing the bottom, top and side of the cylinder and validate if they are on the surface of the cylinder itself (rather than on its extension). 
Out of those validated crossing points we select the two closest ones and the difference between their distances is $x'$, the total distance inside the cylinder measured along the direction specified by $\phi_p$.

\begin{figure}
    \centering
    \includegraphics[width=0.99\linewidth]{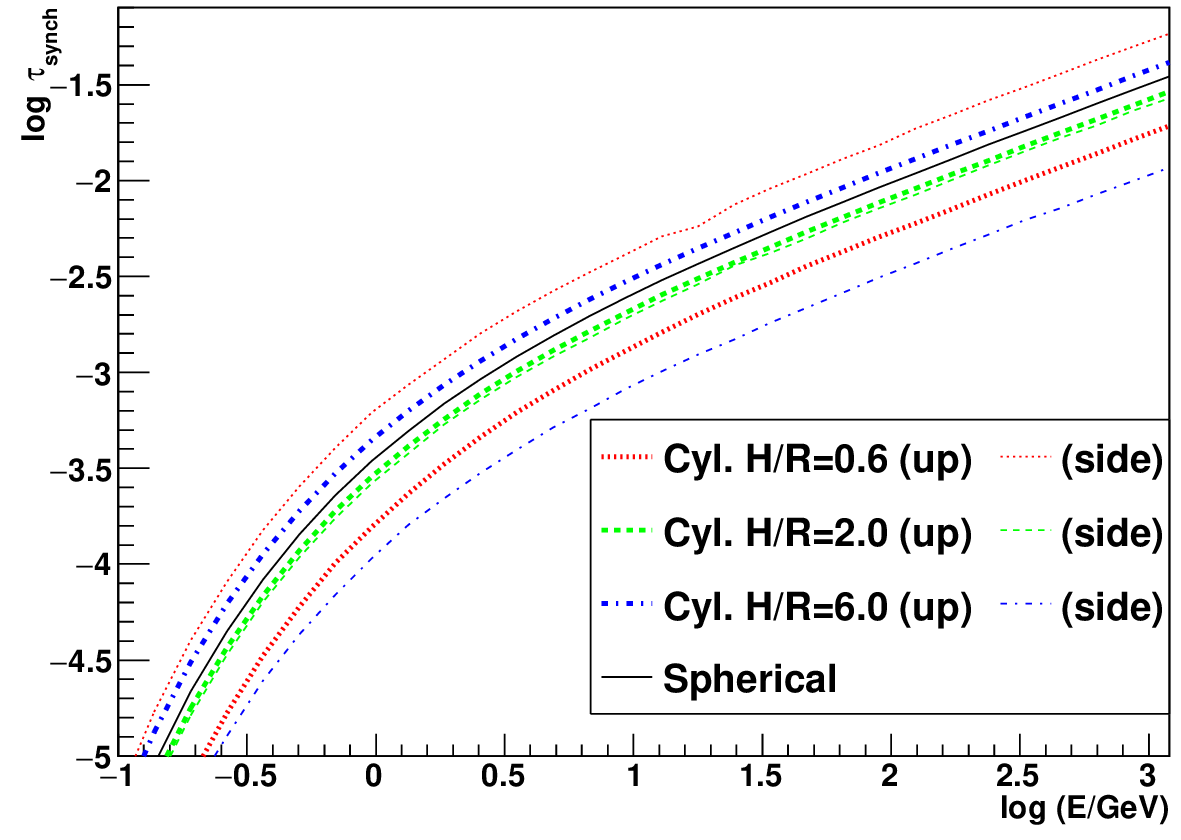}
    \caption{Optical depth for gamma rays produced in the center of the emission region. Black solid line: the case of spherical emission region. 
    Dotted (red), dashed(green) and dot-dashed(blue) line are for cylinders with H/R ratio of 0.6, 2 and 6 respectively.
    Thick lines show the movement of the gamma-ray upwards (along the axis of the cylinder, $\mu_o=1$), while thin lines are for gamma rays escaping to the side ($\mu_o=0$).
    The emission region parameters are as in Fig.~\ref{fig:abs}.}
    \label{fig:abs_dir_cyls}
\end{figure}

Example calculations for the optical depth are shown in Fig.~\ref{fig:abs_dir_cyls}
The absorption of gamma rays emitted in the center of the emission region depends on both the H/R ratio and the observation angle. 
In case of elongated cylinders it is strongest along the axis of the cylinder, while for flat cylinders it is strongest towards the side of the cylinder. 

In order to evaluate the average absorption from the whole emission region we follow the same approach as in Eq.~\ref{eq:a_tot}. 
Due to different symmetry the averaging needs to be done over $r_e$, $\phi_e$, $h_e$ and the result will also depend on the observation angle

\begin{equation}
    A_{\rm cyl}=\frac{\int_0^1 dr_e r_e \int_0^{2\pi} d\phi_e \int_0^H dh_e F_c(r_e, \phi_e, h_e, \mu_o) e^{-\tau}} 
    {\int_0^1 dr_e r_e \int_0^{2\pi} d\phi_e \int_0^H dh_e F_c(r_e, \phi_e, h_e, \mu_o)},\label{eq:a_cyl}
\end{equation}
where $F_c$ is the emission from a particular location in the emission region towards direction of the observer. 

\begin{figure*}
    \centering
    \includegraphics[width=0.33\linewidth]{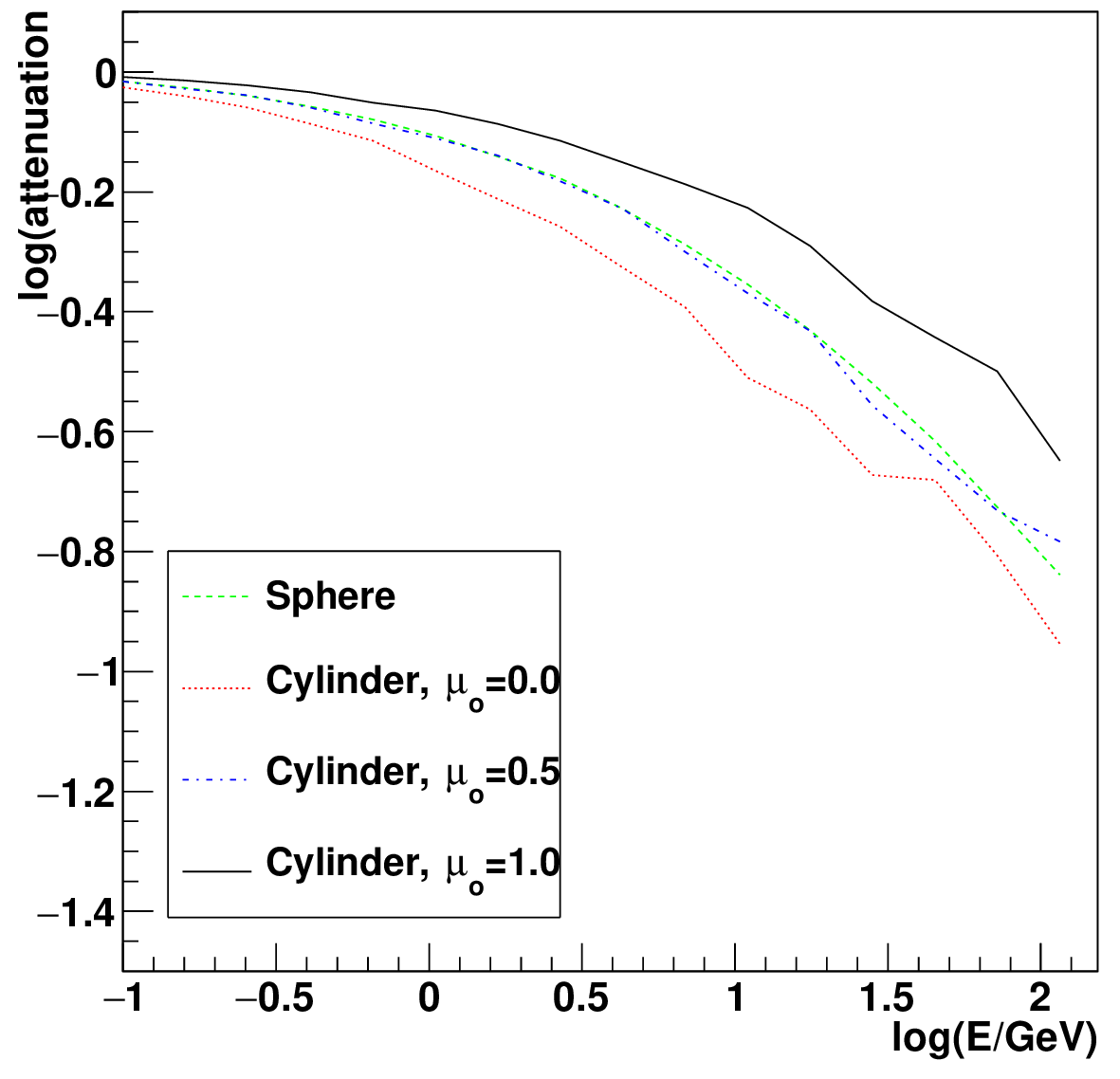}
    \includegraphics[width=0.33\linewidth]{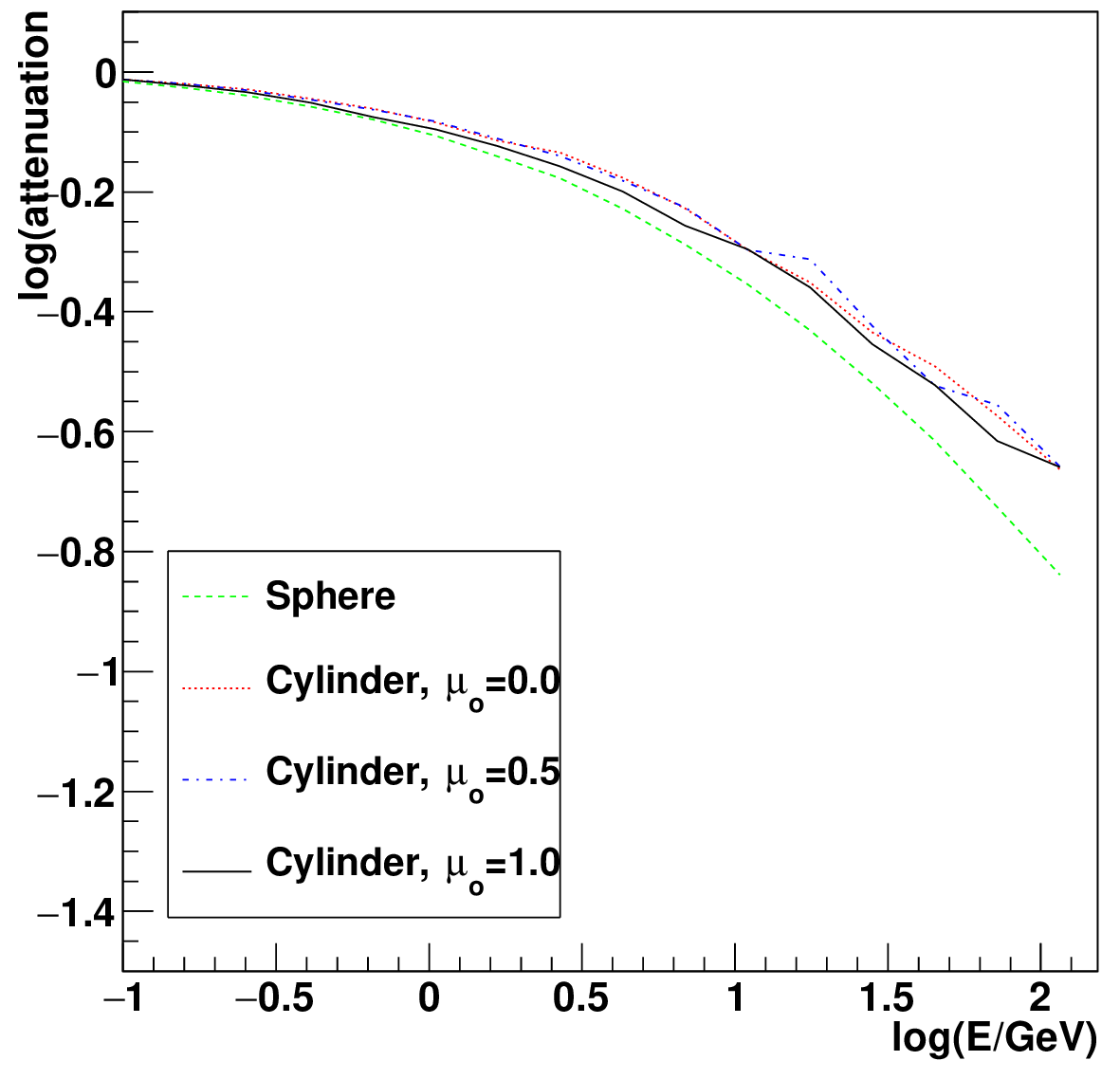}
    \includegraphics[width=0.33\linewidth]{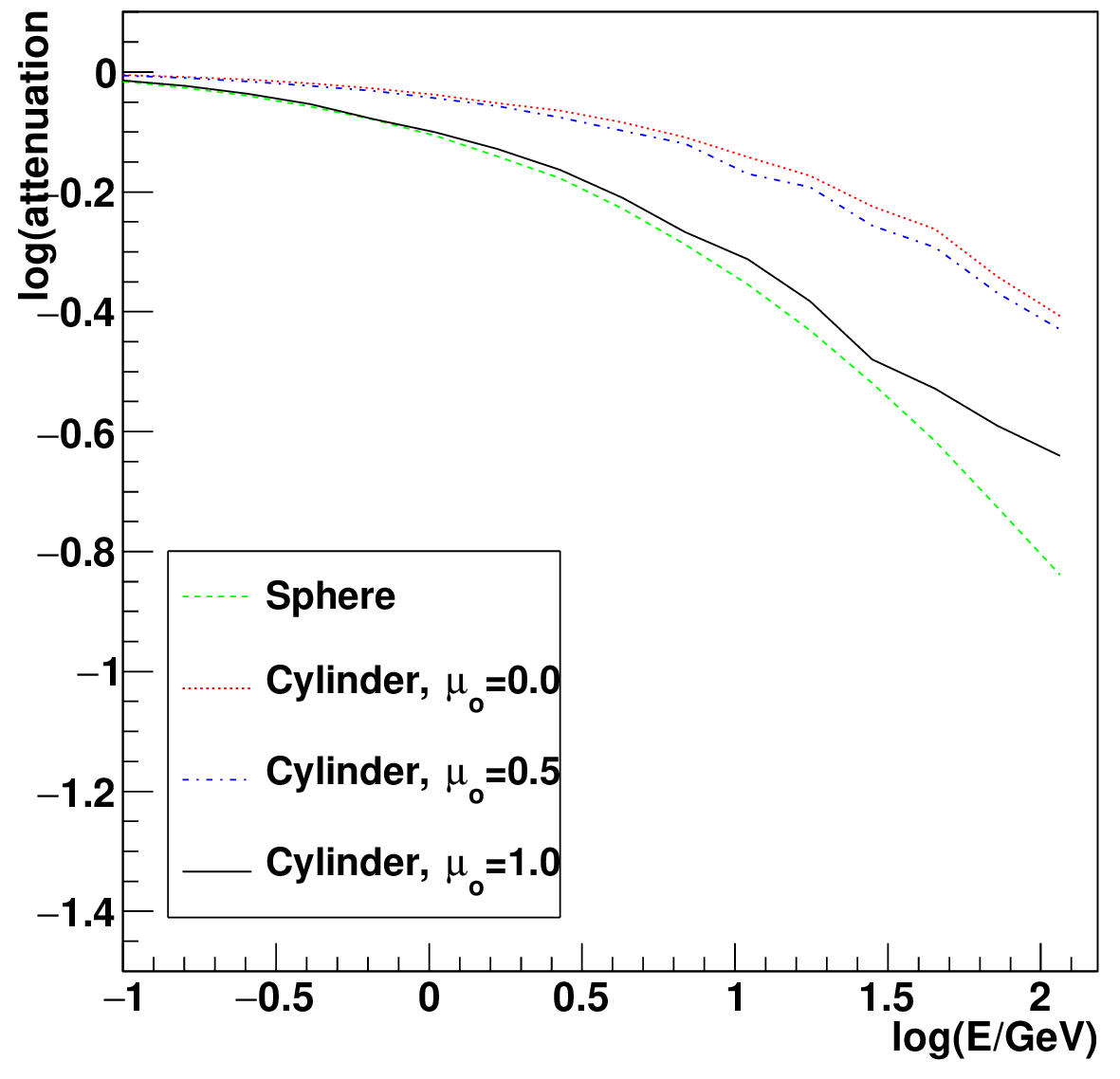}
    \caption{Comparison of the attenuation factors for cylindrical emission regions with H/R ratio of 0.6 (left panel), 2 (middle) and 6 (right panel) with the one of a spherical blob (green dashed line). 
    Cosine of the observation angle with respect to the axis of the cylinder is $\mu_o=0$ (red dotted) 0.5 (blue dot-dashed) and 1 (black). 
    The emission region parameters are as in Fig.~\ref{fig:att_comp}}
    \label{fig:abs_cyl}
\end{figure*}

The example calculations of the absorption factors are shown in Fig.~\ref{fig:abs_cyl}.
The underlying integrals are highly dimensional: the averaging is done by three-dimensional integral from a multiplication of $F_c$ factor and corresponding absorption. 
The $F_c$ factor is a four-dimensional integral, but it is computed as a two-dimensional integral from a tabularized two-dimensional integral. 
The absorption in turn is also a four-dimensional integral. 
Because of this, small numerical instabilities can be observed as $\sim10\%$ point-to-point jumps in this figure. 
The obtained results corroborate those shown in Fig.~\ref{fig:abs_dir_cyls}.
Geometrically the $H/R=2$ cylinder is the closest one to the sphere, resulting also with a similar absorption factors, that are nearly independent on the observation angle. 
In the case of an elongated cylinder the absorption is the most pronounced in the direction of the axis. 
Similarly, in the case of a flat cylinder the strongest absorption occurs if it is observed from the side. 
\section{Conclusions}

Using numerical calculations we investigated the effects of the inhomogenuity and anisotropy of the radiation field and of the emission region (i.e. of the electron distribution) in the SSC scenario for electrons isotropized by the magnetic field.
Using the applied methodology, we reproduced the scaling factor of $3/4$ on the obtained IC emission with respect to homogeneous and isotropic radiation field. 
We also investigated a more realistic scenario or a smoothly distributed electron distribution following a 3D Gaussian. 
We showed that despite anisotropy of the radiation field, in the case of isotropic electrons and spherical symmetry of the emission region the resulting IC emission is also modified just by a global scaling factor.
Such a factor derived for 3D Gaussian distribution is computed numerically as 0.222. 
We also considered a non-spherically symmetric emission region. 
Surprisingly, even in this case we found the emission to be isotropic.
However, cylinders with a different height to radius ratios normalized to have the same synchrotron emission will have IC emission strongly dependent on this ratio, with flat cylinders being more efficient in IC production than narrow ones. 
Also in this case the emission can be computed with regular spherical blob formulae, applying a simple scaling factor, that depends on this ratio.
While all the calculations were done in the reference frame of the emission region, the typical relativistic flux transformations can be used to apply them to beamed sources such as blazars.

The fact that for a broad range of emission region shape assumptions, as long as the electrons are isotropic, the resulting IC emission in the SSC process can be calculated simply from homogeneous spherical blob scenario corrected with a global factor is reassuring for commonly used modelling tools such as agnpy or JetSet.
It means that even while the top-hat spatial distribution of electrons in spherical blob scenario assumed in those tools is a crude approximation of the reality, more complex and more realistic models can still be reproduced by simple scaling factors.
Because of the degeneracy of the classical SSC model with its input parameters (see e.g. \citealp{2017A&A...603A..31A}), those scaling factors can also be achieved with a combination of scaling of individual parameters of the model (such as emission region radius and normalization of the electron distribution).
This allows precise calculations without excessively complicated multidimensional integrals that are computationally expensive and would significantly hinder application to fitting of the data. 

We also investigated how these effects affect the absorption of the produced SSC radiation. 
In this case the situation is much more complex due to an interplay of the emission and absorption efficiency in different parts and for different directions in the emission region.
Nevertheless, for the investigated case of the spherical blob, as long as the absorption is not dramatically strong, the full calculations results are close to the simple approximation of homogeneously emitting slab of material. 
This again confirms the robustness of this approximation in commonly used model fitting tools. 
We also showed that in the case of more realistic, 3D Gaussian distribution of electrons such absorption effects are much smaller. 
This is also the case (but to a smaller degree) if the 3D Gaussian distribution is compactified to reproduce the same level of synchrotron and IC emission as a spherical blob.
In the case of the cylindrical emission region, the absorption breaks the observed independence of the flux on the observation angle. 
Elongated and flat cylinders suffer a stronger absorption when observed from the top and side respectively. 

\section*{Data availability statement}
No new data were generated or analysed in support of this research. 
The simulation software  underlying this article will be shared on reasonable request to the corresponding author.

\section*{Acknowledgements}
This research is supported by the grant from the Polish National Science Centre No. 2023/50/A/ST9/00254.
The author thanks W. Bednarek for his comments and scientific discussions on the manuscript. 
The author would also like to thank anonymous reviewer for the comments that helped to improve the manuscript. 





\begin{thebibliography}{99}
\bibitem[Abe et al.(2024)]{2024MNRAS.535.1484A} Abe, S., Abhir, J., Abhishek, A., et al.\ 2024, \mnras, Constraints on VHE gamma-ray emission of flat spectrum radio quasars with the MAGIC telescopes, 535, 2, 1484. doi:10.1093/mnras/stae2313
\bibitem[Aharonian \& Atoyan(1981)]{aa81} Aharonian, F.~A. \& Atoyan, A.~M.\ 1981, \apss, 79, 321. doi:10.1007/BF00649428
\bibitem[Ahnen et al.(2017)]{2017A&A...603A..31A} Ahnen, M.~L., Ansoldi, S., Antonelli, L.~A., et al.\ 2017, \aap, Multiband variability studies and novel broadband SED modeling of Mrk 501 in 2009, 603, A31. doi:10.1051/0004-6361/201629540
\bibitem[Alves Batista et al.(2023)]{2023arXiv231200409A} Alves Batista, R., Amelino-Camelia, G., Boncioli, D., et al.\ 2023, Class. Quantum Grav. 42 032001,2025; doi:10.1088/1361-6382/ad605a 
\bibitem[Chen et al.(2011)]{2011MNRAS.416.2368C} Chen, X., Fossati, G., Liang, E.~P., et al.\ 2011, \mnras, 416, 3, 2368. doi:10.1111/j.1365-2966.2011.19215.x
\bibitem[Crusius \& Schlickeiser(1986)]{cs86} Crusius, A. \& Schlickeiser, R.\ 1986, \aap, 164, L16
\bibitem[Dermer et al.(2009)]{2009ApJ...692...32D} Dermer, C.~D., Finke, J.~D., Krug, H., et al.\ 2009, \apj, 692, 1, 32. doi:10.1088/0004-637X/692/1/32
\bibitem[Finke et al.(2008)]{fi08} Finke, J.~D., Dermer, C.~D., \& B{\"o}ttcher, M.\ 2008, \apj, 686, 181. doi:10.1086/590900
\bibitem[Gould \& Schr{\'e}der(1967)]{1967PhRv..155.1404G} Gould, R.~J. \& Schr{\'e}der, G.~P.\ 1967, Physical Review, 155, 5, 1404. doi:10.1103/PhysRev.155.1404
\bibitem[Gould(1979)]{1979A&A....76..306G} Gould, R.~J.\ 1979, \aap, 76, 3, 306. 
\bibitem[Gulati et al.(2024)]{2024ApJ...977....9G} Gulati, S., Bhattacharya, D., \& Sreekumar, P.\ 2024, \apj, Constraining the Location of the {\ensuremath{\gamma}}-Ray Emission Region in Radio-loud AGN 3C 380, 977, 1, 9. doi:10.3847/1538-4357/ad891e
\bibitem[Hahn(2015)]{ha15} Hahn, J. 2015, in International Cosmic Ray Conference, Vol. 34, 34th International Cosmic Ray Conference (ICRC2015), 917
\bibitem[Li(2022)]{2022JCAP...02..025L} Li, H.-J.\ 2022, \jcap, Relevance of VHE blazar spectra models with axion-like particles, 2022, 2, 025. doi:10.1088/1475-7516/2022/02/025
\bibitem[Mankuzhiyil et al.(2011)]{2011ICRC....8..167A} Mankuzhiyil, N., Ansoldi, S., De Caneva, G., et al.\ 2011, International Cosmic Ray Conference, 8, 167. doi:10.7529/ICRC2011/V08/1158
\bibitem[Maraschi et al.(1992)]{ma92} Maraschi, L., Ghisellini, G., \& Celotti, A.\ 1992, \apjl, 397, L5. doi:10.1086/186531
\bibitem[Moderski et al.(2005)]{mo05} Moderski, R., Sikora, M., Coppi, P.~S., et al.\ 2005, \mnras, 363, 954. doi:10.1111/j.1365-2966.2005.09494.x
\bibitem[Nigro et al.(2022)]{2022A&A...660A..18N} Nigro C., Sitarek J., Gliwny P., Sanchez D., Tramacere A., Craig M., 2022, A\&A, 660, A18. doi:10.1051/0004-6361/202142000
\bibitem[Sol \& Zech(2022)]{2022Galax..10..105S} Sol, H. \& Zech, A.\ 2022, Galaxies, Blazars at Very High Energies: Emission Modelling, 10, 6, 105. doi:10.3390/galaxies10060105
\bibitem[Tramacere et al.(2011)]{2011ApJ...739...66T} Tramacere, A., Massaro, E., \& Taylor, A.~M.\ 2011, \apj, 739, 66. doi:10.1088/0004-637X/739/2/66
\bibitem[Tramacere(2020)]{2020ascl.soft09001T} Tramacere, A.\ 2020, Astrophysics Source Code Library. ascl:2009.001
\bibitem[Tramacere et al.(2022)]{2022A&A...658A.173T} Tramacere, A., Sliusar, V., Walter, R., et al.\ 2022, \aap, 658, A173. doi:10.1051/0004-6361/202142003
\bibitem[Zhang et al.(2024)]{2024ApJ...967...93Z} Zhang, H., B{\"o}ttcher, M., \& Liodakis, I.\ 2024, \apj, 967, 2, 93. doi:10.3847/1538-4357/ad4112
\end{thebibliography}




\appendix
\section{Analytical calculations}
\subsection{Narrow cylinder in the Thomson regime} \label{sec:cylgeom}
In order to show that for homogeneous, but highly non-spherically symmetric blob the emission does not depend on the observation angle, let us consider a cylinder with height H, and radius much smaller than H. 
It is observed at the angle to the axis, which cosine is $\mu_o$.
In this case, the radiation field will depend only on the height measured along the cylinder, $z$.
Moreover, it can be considered as the sum of the radiation emitted at heights above $z$, i.e. from a part of the cylinder with height $h=H-z$ and below $h=z$ (from a part of the cylinder with height $z$).
For homogeneous emission of the soft radiation, such density of radiation coming from the top/bottom would only depend then on $h$ and thus can be expressed as $F(h)$.
Due to diverging integrals for an infinitesimally thin cylinder the asymptotic value of $F(h)$ cannot be computed; however we can use it to show the independency of the emission on the observation angle. 
The total inverse Compton emission of the cylinder will be proportional to integral over the height of the blob from the sum of radiation from the top part and the bottom, however weighted with the different interaction factors. 
In the assumption of the Thomson region, where the interaction cross section does not depend on the energy of the soft photon, the interaction factor will be given by $1-\mu_s$, where $\mu_s$ is the cosine of the scattering angle.
As the cylinder is considered to be thin, all the soft photons would be propagating along its height.
Therefore $\mu_s$ is equal to $\mu_o$ for the soft photons coming from below of the electron and $-\mu_o$ for the photons from the above.
In summary the integral to which the emission should be proportional can be written as:
\begin{eqnarray}  
    I&=&\int_0^H dz \left(F(H-z)(1+\mu_o) + F(z)(1-\mu_o)\right) = \nonumber\\
    &=& \int_0^H dz\left(F(H-z)+F(z)\right) \nonumber\\
    &+&\mu_o\int_0^H dz\left(F(H-z) - F(z)\right).\label{eq:A1}
\end{eqnarray}
Since $\int_0^H dz F(H-z) = \int_0^H dz F(z)$, for any functional form of $F(z)$ the second integral vanishes and $I=2\int_0^H dz F(z)$, which is independent of $\mu_o$. 

\begin{figure}
    \centering
    \includegraphics[trim = 50 0 50 0 , clip,width=0.5\linewidth]{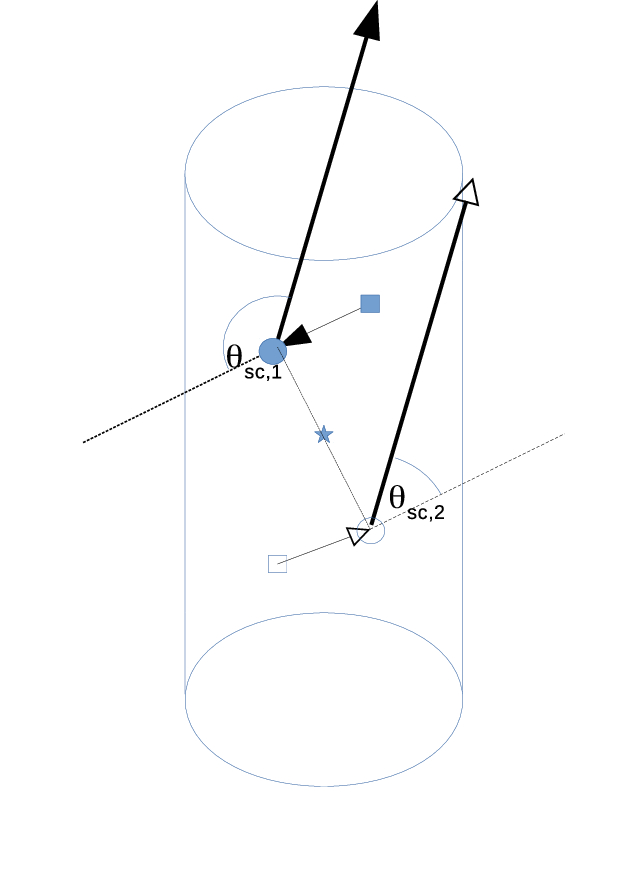}
    \caption{Geometry of interactions in cylindrical emission region. Thick arrow shows the emission at which a gamma ray is emitted, while the thin arrow shows the path of the scattered soft photon and the interaction point is marked with a circle. Two cases with positions reflected with respect to the centre of the emission region (star) are shown with filled and empty arrows and markers.
    The corresponding interaction angles are $\theta_{sc,1}$ and $\theta_{sc,2}$.}
    \label{fig:cylinder}
\end{figure}

A similar argument can be applied to any spatial distribution of isotropic electrons, as long as their density distribution has central symmetry (see Fig.~\ref{fig:cylinder}). 
For any observation angle, and any combination of the position of soft photon source and interaction point one can consider a point reflection with the same density of soft photons and scattering electrons. 
The scattering angles are then related by $\theta_{sc,1}=\pi-\theta_{sc,2}$. 
Therefore, similarly to Eq.~\ref{eq:A1} the parts of the integral linear in the cosine of the scattering angle will cancel out when integrating over the whole emission region.




\bsp	
\label{lastpage}

\end{document}